\documentclass[twocolumn]{aastex701}

\usepackage{amsmath}
\usepackage[normalem]{ulem}
\usepackage{xcolor}
\usepackage{lineno}

\makeatletter
 
\newcommand{\Rmnum}[1]{\expandafter\@slowromancap\romannumeral #1@}

\usepackage{atbegshi}
\newcommand\showtimer{%
  \message{^^Jtimer: \the\numexpr\the\pdfelapsedtime*1000/65536\relax}%
  \pdfresettimer}
\AtBeginDocument{\showtimer}
\AtBeginShipout {\showtimer}

\def\be{\begin{equation}}
\def\ee{\end{equation}}
\def\bea{\begin{eqnarray}}
\def\eea{\end{eqnarray}}
\def\ba{\begin{array}}
\def\ea{\end{array}}

\def\nn{\nonumber}

\def\teal{\textcolor{teal}}

\makeatother

\begin{document}

\linenumbers

\title{Morphological complexity of NGC 628 - a multiwavelength multiscale analysis \\
using the ordinal pattern framework}

\author[orcid=0000-0002-5071-7371]{Athokpam Langlen Chanu}
\affiliation{Asia Pacific Center for Theoretical Physics, Pohang, 37673, Republic of Korea}
\affiliation{Department of Physics, Pohang University of Science and Technology (POSTECH), Pohang, 37673, Republic of Korea}
\email{athokpam.chanu@apctp.org}  

\author[orcid=0009-0005-6072-9252]{S Amrutha} 

\affiliation{Indian Institute of Astrophysics, Koramangala II Block,  
  Bangalore 560034, India}
  \affiliation{
Pondicherry University, R.V. Nagar, Kalapet, 605014, Puducherry, India}
\email{amrutharao99@gmail.com}

\author[orcid=0000-0002-7385-8273]{Pravabati Chingangbam}
\affiliation{Indian Institute of Astrophysics, Koramangala II Block,  
  Bangalore 560034, India}
  \affiliation{
Pondicherry University, R.V. Nagar, Kalapet, 605014, Puducherry, India}
\email[show]{prava@iiap.res.in (Corresponding author)}

\author[orcid=0000-0001-9521-6397]{Changbom Park}

\affiliation{School of Physics, Korea Institute for Advanced Study, 85 Hoegiro, Dongdaemun-gu, Seoul, Republic of Korea 02455}
\email{cbp@kias.re.kr}

\begin{abstract}
As statistical systems, galaxies exhibit a rich interplay between organized structure and stochastic fluctuations across a broad range of spatial scales. This duality motivates the need for quantitative frameworks capable of capturing their morphological complexity. The ordinal patterns framework, along with its associated statistical measures: permutation entropy ($H$), disequilibrium ($D_E$), statistical complexity ($C$), and ordinal network node entropy, has recently emerged as a powerful tool for analyzing such complexity in physical systems. 
We apply this framework in a multiwavelength, multiscale analysis of the galaxy NGC 628, utilizing observations in the near-ultraviolet, near-infrared, mid-infrared, and millimeter bands. Our results reveal a characteristic spatial scale of approximately 200 parsecs, marking the transition from small-scale structures influenced by star formation and stellar feedback to larger-scale morphology governed by the galaxy's dynamics. 
Furthermore, we find that the $C$ vs. $H$ trajectories for all wavelengths converge toward a common attractor curve, consistent with the behavior of isotropic Gaussian random fields. This convergence suggests a universal statistical behavior in galactic structure at large scales, despite the differing physical processes traced by each wavelength.

\end{abstract}

\keywords{\uat{Galaxies}{573}, \uat{Spiral galaxies}{1560}, \uat{Galaxy structure}{622}, \uat{Star formation regions}{1565}, \uat{Interstellar medium}{847}}

\section{Introduction} \label{sec:s1}

Complexity of statistical systems is a multifaceted concept, with numerous complexity measures proposed across disciplines~\citep{lloyd2001measures,friedrich2011approaching,roli2019complexity}. Traditional complexity measures include Lyapunov exponents~\citep{wolf1985determining}, fractal dimensions~\citep{mandelbrot1985self}, and entropy~\citep{cover1999elements}, each offering a distinct perspective on the underlying characteristics of complex systems. 

~\cite{lopez1995statistical} introduced a statistical measure of complexity based on the distance of the probability distribution of `accessible states' of a given system from an equiprobable distribution, multiplied with the Shannon entropy~\citep{shannon1948mathematical}. This complexity measure and Shannon entropy encode complementary information of order and disorder. The so-called {\em complexity-entropy plane} then serves as a phase space on which different physical systems take values,  allowing for distinction by their degree of complexity. This characterization of physical systems depends on the `scale of observations' as the space of accessible states varies with observation scale. 

~\cite{bandt2002permutation} introduced the so-called {\em permutation entropy} ($H$), as an information-theoretic measure of the complexity of one-dimensional empirical data, by constructing accessible states based on comparison of neighboring data values and symbolic dynamics~\citep{balasis2013statistical}. ~\cite{rosso2007distinguishing} extended the formalism and proposed a statistical complexity measure ($C$) based on the 
notion of complexity introduced in~\cite{lopez1995statistical}, and used the complexity-entropy ($CH$) plane to distinguish different systems. ~\cite{ribeiro2012complexity} and ~\cite{zunino2016discriminating} further extended the formalism to two-dimensional (2D) data, which is used to investigate spatial structures~\citep{zunino2016discriminating,sigaki2018history,sigaki2019estimating,morel2021multiscale} in image analysis. The formalism of $CH$-plane has been widely used to distinguish {\em complex structures} across different physical domains~\citep{mao2019multivariate,chanu2024exploring,chanu2025,singh2026hybrid}. Further, one can construct the so-called {\em ordinal networks} from transitions among accessible states~\citep{pessa2019characterizing} that can extract additional distinguishing features in different datasets (see e.g.,~\cite{pessa2020mapping,pessa2021ordpy}).

Astrophysical systems - ranging from sunspot cycles, turbulent plasmas, orbital systems to galaxy morphologies and large-scale structure - exhibit rich complexity shaped by nonlinear processes, feedback, and multiscale interactions~\citep{ shore2003galaxies,regev2006chaos,consolini2009complexity,aschwanden2011self, fiore2024dynamical,langlen2024analysis,hyman2025peccary}. This paper is focused on the complexity of galaxy morphologies. Galaxies constitute an important dynamical component of our observable universe. Various studies have shown that the morphology of galaxies correlates with their physical properties, including mass, specific star formation rate (sSFR), gas fraction, and integrated colors (e.g., (g–r), (B–V)), among others~\citep{1976ApJ...204..649G, Poggianti_2009, 2019MNRAS.488.3929C}. For instance, elliptical galaxies are generally more massive than spiral galaxies. Since elliptical galaxies have already consumed most of their gas to form stars, they exhibit low gas fractions and show no recent star formation, making them appear red compared to late-type spirals. Thus, morphological classification is crucial for understanding galaxy formation, evolution, and the underlying physical processes.

Existing classification methods primarily rely on visually distinguishing morphological features~\citep{1936rene.book.....H, DeVaucouleurs1959}. More recent approaches use improved photometric and spectroscopic data \citep{kauffman1993origins, 1995AJ....110.1071C, 2004AJ....128..585Y}, and automated intelligent techniques, including machine learning, to classify large numbers of galaxies~\citep{2011MNRAS.410..166L, 2019Ap&SS.364...55Z, 2023ApJ...942L..42R}. There are other non-parametric techniques describing the shape and light distribution of galaxies which are commonly used to classify the galaxies \citep{2004AJ....128..163L, 2005PASA...22..118G, Conselice}. 

The structural and dynamical features of galaxies make them ideal for analysis using tools from the statistical physics of complex systems. In this pilot study, we assess the use of ordinal patterns and associated measures - permutation entropy, statistical complexity, and ordinal network node entropy - for classifying galaxy morphology. We select NGC 628 (the Phantom Galaxy), a well-studied, face-on spiral galaxy with high-resolution, multiwavelength data, meeting criteria for 2D-analysis, emission diversity, and data accessibility. NGC 628 has been extensively studied across a broad spectrum of wavelengths, from radio frequencies to X-rays (see refs ~\cite{1992A&A...253..335K, 2002ApJ...572L..33S, 2018ApJ...863L..21K, Lee_2023}). In this study, using near-ultraviolet (NUV), near-infrared (NIR), mid-infrared (MIR), and millimeter (mm) images, we quantify its morphological complexity across wavelengths.  

The ordinal pattern framework and in particular the $CH$-plane has been previously used to study magnetic flux ropes in the corona of the sun~\citep{gekelman2019spiky}, turbulence in the solar wind~\citep{weck2015permutation,weygand2019jensen}, electron plasma pressure filaments (~\cite{karbashewski2022magnetized}), characterization of exoplanet systems~\citep{gilbert2020information,bartlett2022assessing}, x-ray ~\citep{lovallo2011complexity} and radio astrophysical sources~\citep{segal2019identifying}. Applying information-theoretic metrics such as entropy and statistical complexity has yielded valuable insights across astrophysical systems. For example, \cite{bartlett2022assessing} demonstrates that these metrics effectively characterize planetary features and can assist in detecting potential biospheres in exoplanet studies. \cite{weygand2019jensen} uses them to reveal the stochastic behavior of solar wind structures, including interplanetary coronal
mass ejections (ICMEs), co-rotating interaction regions (CIRs), and turbulent magnetic fluctuation intervals. Likewise, \cite{segal2019identifying} shows their ability to distinguish simple and complex radio galaxy morphologies in data from the Australia Telescope Large Area Survey.  

This paper is organized as follows. Section~\ref{sec:s2} outlines the framework of ordinal patterns, permutation entropy, statistical complexity, and ordinal networks, along with their physical interpretations. Section~\ref{sec:s3} describes the multiwavelength data and physics of NGC 628. 
In section~\ref{sec:s4}, we present our main analysis and results. Finally, section~\ref{sec:s5} concludes with a summary and discusses the implications of our findings and future directions.

\section{Framework of ordinal patterns, permutation entropy, statistical complexity, and ordinal networks} 
\label{sec:s2}

This section introduces the main concepts used in our analysis, namely, ordinal patterns, permutation entropy, statistical complexity, complexity-entropy causality plane, 
and ordinal networks. This section follows~\cite{pessa2021ordpy}.

Consider an  $N_x\times N_y$ data matrix $Y$ whose elements are $Y^{j}_i$, with $i=1,2,\ldots,N_x,\ j=1,2,\ldots,N_y$. For a fixed $(i,j)$, let $y$ be a sub-matrix of size $d_x\times d_y$: 
   \begin{equation} y=  
\begin{bmatrix}
Y_{i}^{j} & Y^{j+1}_{i} & \dots & Y^{j+(d_y-1)}_i \\
Y_{i+1}^{j} & Y^{j+1}_{i+1} & \dots & Y^{j+(d_y-1)}_{i+1} \\
\vdots &\vdots &\ddots &\vdots \\
Y_{i+(d_x-1)}^{j} & Y^{j+1}_{i+(d_x-1)} & \dots & Y^{j+(d_y-1)}_{i+(d_x-1)}
\end{bmatrix}.
 \end{equation}
$d_x,d_y$ are referred to as `embedding dimensions'. The smallest choice is $d_x=2=d_y$, in which case $y$ is a $2\times 2$ matrix. 
Moreover, they are chosen to satisfy $d_x\ll N_x,\,d_y\ll N_y$ with $(d_xd_y)! \ll N_xN_y$. This condition is necessary to get a reliable estimate of the ordinal probability distribution, which will be defined shortly.

Let us denote $n_x\equiv N_x-(d_x-1)$ and $n_y\equiv N_y-(d_y-1)$.  By varying $i$ from 1 to $n_x$, and $j$ from 1 to $n_y$ in steps of one, we can construct $n_x n_y$ number of such $y$ sub-matrices. In general, the steps for $i$ and $j$ can be arbitrary positive integers, usually known as {\em embedding delays}. Here we restrict to unit embedding delays.  

{\bf{\em Permutation state or ordinal pattern}}: Every sub-matrix $y$ can be mapped to a `permutation state' or `ordinal pattern' by the following procedure. 
\begin{enumerate}
\item {\em Flatten the elements of $y$}: Write the elements of $y$ as a one-dimensional sequence denoted by $z=\{ z_1,z_2,\dots,z_{d_xd_y}\}$ (referred to as the `flattening' step). 
There is no unique way to flatten $y$. We choose one way of doing it consistently for all $y$.
\item {\em Map $z$ to a symbolic sequence $A$}: We assign an integer sequentially to each element of $z$ as, 
\be
\{z_m\}\to \{(m-1)\} \equiv A,\ \ m=1,2,\dots,{d_xd_y}.
\ee
$A$ is a symbolic sequence because its integer elements only serve as position markers or indices for the elements of $z$. 
\item {\em Sorting and permutation}: Next, the elements of $z$ are sorted in ascending order to get another sequence $z_S$, resulting in the elements of $A$ getting shuffled or permuted to get another sequence $A_S$. 
The new sequence $A_S$ is referred to as a `permutation state' or `ordinal pattern'. Other terms such as `permutation pattern', `ordinal state', and `permutation symbol' are also used in the literature. 
\end{enumerate}
The steps 1-3 above map every sub-matrix $y$ to a permutation state. We introduce the symbol $\pi_i^j, \,i=1,2,\dots,n_x,j=1,3,\dots,n_y$, to denote a permutation state. The sequence  $\{\pi_i^j\}_{i=1,2,3,\dots,n_x}^{j=1,2,3,\dots,n_y}$ is referred to as the {\em ordinal sequence}. It is just the set of all $n_x n_y$ number of permutation states obtained from $Y$. An illustration of the computation of permutation states for an example two-dimensional array is given in the Appendix.

{\bf{\em Ordinal probability distribution}}: Given $d_xd_y$  number of distinct non-negative integers we have $(d_xd_y)!$ distinct states. Under the condition  $N_xN_y \sim n_xn_y \gg (d_xd_y)!$, not all states in the ordinal sequence 
will be distinct. 
 Let us denote each distinct state by $\psi_k$, where $k=1,2,\ldots,(d_xd_y)!$. Let $n_k$ denote the number of occurrences of each $\psi_k$ in the ordinal sequence. Then the probability of occurrence of each $\psi_k$ is given by:
\begin{equation}
    \rho_k(\psi_k)= \frac{n_k(\psi_k)}{n_xn_y}. 
    \label{eq:rho}
\end{equation}
The function $P=\{\rho_k(\psi_k)\}$ is referred to as the {\em ordinal probability distribution}. It satisfies the normalization condition:
\begin{equation}
\sum_{k=1}^{(d_xd_y)!} P = \sum_{k=1}^{(d_xd_y)!} \rho_k(\psi_k) = 1.
\end{equation}
Having introduced the basic concepts of ordinal patterns, we now introduce the statistical quantities that we will use to analyze the galaxy data.

\subsection{Permutation Entropy, Statistical Complexity, and the Complexity-Entropy (\texorpdfstring{$CH$}{CH}) causality plane} 
\label{sec:s2a}

\noindent{\bf{\em Permutation entropy, $H[P]$}}: The functional $S[P]$ expressed by:
\begin{equation}
    S[P]=-\sum_{k=1}^{(d_xd_y)!} \rho_k(\psi_k) \log_2\rho_k(\psi_k), \label{eq:pe1}
\end{equation}
is the entropy of the permutation states~\citep{bandt2002permutation}. Then the {\em normalized} permutation entropy, $H[P]$~\citep{rosso2007distinguishing}, is defined as,
\begin{equation}
    H=\frac{S[P]}{\log_2 (d_xd_y)!}, \label{eq:npe}
\end{equation}
where the normalization factor $\log_2 (d_xd_y)!$ is the maximum entropy, thus $H$ satisfies $0 \leq H\leq 1$. $H$ (or $S$) provides a measure of the degree of disorder or randomness of the data $Y$, with $H\to 1$ indicating high disorder. 

 $H$ is exactly one when $Y$ consists of totally random elements and all distinct permutation states are equiprobable, whereas when physical correlations are present in the data across scales set by $(d_x,d_y)$, some states have higher probabilities, hence $H < 1$. The higher the correlation strength, the lower the value of $H$.  

\vskip .4cm
\noindent{\bf{\em Statistical complexity, $C[P,U]$}}:  Let $U$ denote the uniform probability distribution, $U=\{u_k\}$, with $k=1,2,\dots,(d_xd_y)!$, where $u_k=1/(d_xd_y)!$. The Jensen-Shannon divergence between the probability distributions $P$ and $U$ is given by,
\be
D[P,U] = S\bigg[\frac{(P+U)}{2}\bigg]-\frac{S[P]}{2}-\frac{S[U]}{2}, \label{eq:d}
\ee
with permutation entropy of $P$, $S[P]$ (eq.~\ref{eq:pe1}), and permutation entropy of $U$, $S[U]$ given by,
\be
    S[U]=-\sum_{k=1}^{(d_xd_y)!} u_k \log_2 u_k. \label{eq:PU}
\ee
$P+U$ is the mixture distribution of $P$ and $U$, and its corresponding permutation entropy is 
\be
S\bigg[\frac{P+U}{2}\bigg]  =  -\sum_{k=1}^{(d_xd_y)!} \left(\frac{\rho_k + u_k}{2}\right) \log_2 \left(\frac{\rho_k +u_k}{2}\right).
\ee
$D$ of eq.~\eqref{eq:d} measures how different $P$ is from $U$. It is non-zero when there exist more likely states in the space of states constructed from $Y$.

The so-called {\em disequilibrium}~\citep{martin2003statistical}, denoted by $D_E$, is defined as the normalized Jensen-Shannon divergence, 
\be D_E[P,U]=\frac{D[P,U]}{D_{\rm max}}, \ee where $D_{\rm max}$ denote the maximum possible value of $D[P,U]$ obtained when $P$ is one and $U$ vanishes as: 
\bea 
D_{\rm max} &=&-\frac{1}{2}\bigg \{ \left[\frac{(d_xd_y)!+1}{(d_xd_y)!}\right]\log[(d_xd_y)!+1] \nn\\
&& - 2 \log[2(d_xd_y)!] 
+\log (d_xd_y)!\bigg \}.
\eea
$D_E$ (or $D$) provides a measure of order and organization in the data $Y$ since it measures how far the ordinal probability distribution of $Y$ is from an equiprobable distribution (representing equilibrium). By definition, $0\leq D_E\leq 1$.  
Finally, the {\em statistical complexity}, $C$,~\citep{rosso2007distinguishing} is defined to be:
\begin{equation}
    C=D_E[P,U]\ H[P]. \label{eq:scm}
\end{equation}
\vskip .4cm
\noindent{\bf{\em Complexity-Entropy} ($CH$) \em{plane}}: For a given data $Y$, and for chosen $(d_x,d_y)$ values,  the computed values of $H$ and $C$ occupy a point ($H,C$) on the $CH$-plane with $H$ on the $x$-axis and $C$ on the $y$-axis. Datasets of different physical origins, for example, stochastic processes such as white noise, colored noise, and fractional Brownian motion or chaotic processes such as Lorenz, logistic map will typically occupy distinct regions of the $CH$-plane and hence can be distinguished~\citep{rosso2007distinguishing}.  

The entropy and complexity measures defined above have various attractive properties~\citep{bandt2002permutation,rosso2007distinguishing}. The permutation patterns naturally emerge from the data without the need for any parametrization. Moreover, the results are unaffected by the presence of low levels of noise, invariant under monotonic transformations of the data values, and are also computationally inexpensive. 

It is useful to analyze a given 2D-data using different choices of $(d_x,d_y)$ because they probe spatial correlations at varying scales.  For a given  $(N_x, N_y)$, as we increase $d_x$ and $d_y$, the total number of permutation states given by $n_xn_y$ decreases, while the number of distinct states $(d_x,d_y)!$ grows factorially. 
This leads to a decrease of $H$ due to the normalization factor in eq.~\ref{eq:npe}, unless the data contains unusual scale dependence. In comparison, the behavior of $C$ is not straightforward to anticipate. Further, interchanging $d_x$ and $d_y$ is equivalent to rotating the data by 90°, which can yield different $H$ and $C$ if the system lacks rotational symmetry, making such analysis useful for probing anisotropy. 

\subsection{Node Entropies from Ordinal Network}
\label{sec:s2b}

The ordinal patterns from the data matrix $Y$ can be mapped into the nodes of an ordinal network~\citep{small2018ordinal, pessa2019characterizing}. Each basis state $\psi_k$ corresponds to a node, yielding $k=1,2,\dots,(d_xd_y)!$ nodes. Directed edges connect adjacent basis states (horizontal or vertical), representing {\em transition} between the states. Suppose $n_{k\to m}$ denotes the total number of occurrences of a transition $\psi_k \to \psi_m$. Then the {\em weighted adjacency matrix} is:
\begin{equation}
    \rho_{k,m}= \frac{n_{k\to m}}{2n_xn_y-n_x-n_y}. \label{eq:on}
\end{equation}
The denominator represents the total number of horizontal and vertical transitions. Then the normalized transition probability is: 
\be \rho_{k,m}^{'}= \frac{\rho_{k,m}}{ \displaystyle \sum_{m\in \mathcal{O}_k}\rho_{k,m}},
\ee 
where $\mathcal{O}_k$ is the set of all outgoing edges from node $k$. 

We focus on the entropy measures that take into account the probabilistic nature of nodes and edges {\em locally}, and {\em globally} for the ordinal network. The {\em local node entropy}~\citep{mccullough2017multiscale,small2018ordinal,pessa2020mapping}
\begin{equation}
    s_k=-\sum_{m\in \mathcal{O}_k} \rho_{k,m}^{'} \log \rho_{k,m}^{'},
\end{equation}
measures determinism in the transitions at the node level~\citep{pessa2020mapping}; $s_k=0$ if only one edge exits $k$, and is maximum when all edges leaving $k$ have the same weight (equiprobability). 

 The {\em global node entropy}~\citep{pessa2020mapping}
\begin{equation}
    {\rm S}_{\rm gn}=\sum_{k=1}^{(d_xd_y)!} \rho_k s_k, \label{eq:gnentropy}
\end{equation}
is the weighted average of local entropies over all nodes of the network, with $\rho_k$ the probability of occurrence of each $\psi_k$ [eq.~\eqref{eq:rho}].

\subsection{Ordinal pattern analysis for Gaussian random fields - a toy example}
\label{sec:s2c}

\begin{figure*}
\centering
\includegraphics[scale=0.43]{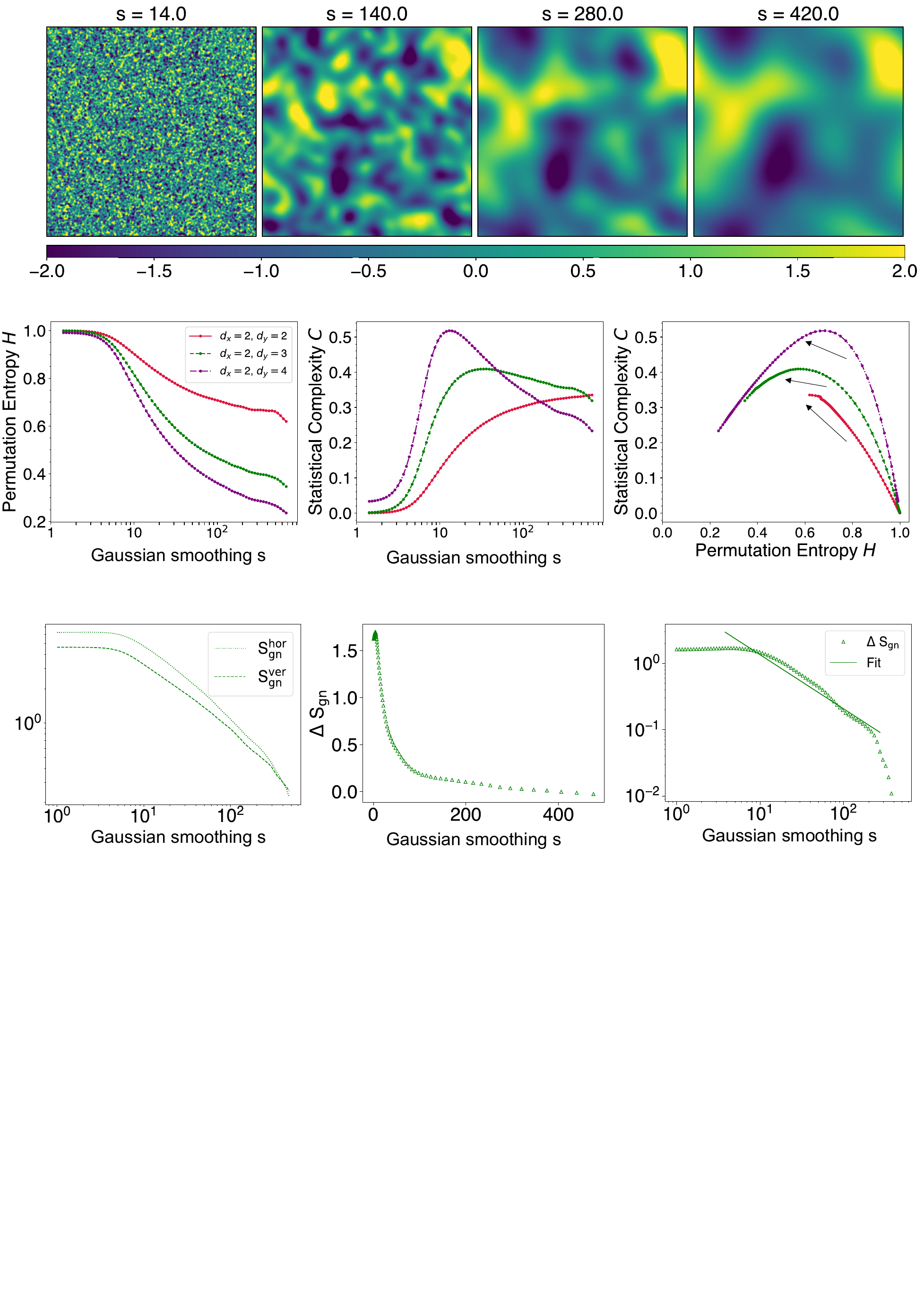}
\vskip -7cm
\caption{{\em Top row}: One realization of a two-dimensional isotropic Gaussian random field generated on a grid of $500^2$ pixels, with increasing smoothing scales (left to right) given in units of pixel numbers. {\em Middle row}: Permutation entropy, $H$, (left)  and statistical complexity, $C$, (middle), as functions of the smoothing scale $s$ at embedding dimensions $(d_x,d_y)=(2,2) \ \mathrm{(red)}, (2,3) \ \mathrm{(green)}, \mathrm{and} (2,4) \ \mathrm{(purple)}$, for the fields in the top row. The right panel shows the $CH$-plane.  The black arrows indicate the direction of increasing $s$. {\em Bottom row}: Horizontal, S$_{\rm gn}^{\rm hor}$, (dotted) and vertical, S$_{\rm gn}^{\rm ver}$, (dashed) global node entropies (left), global node entropy difference, $\Delta$S$_{\rm gn}$, (middle), as functions of the smoothing scale, for  $(d_x,d_y)=(2,3)$. The log-log plot of the right panel shows an inverse power law fit (solid line) to $\Delta$S$_{\rm gn}$ (triangles) as a function of smoothing scale.}
\label{fig:gaussian}
\end{figure*}

 Before analyzing the morphological complexity of NGC 628, we first evaluate permutation entropy ($H$), statistical complexity ($C$), $CH$-plane, and global node entropy (S$_{\textrm{gn}}$), as described in subsections~\ref{sec:s2a} and \ref{sec:s2b}, for realizations of Gaussian random fields (GRFs), which provide benchmarks for interpreting the galaxy's results.

We simulate GRFs with zero mean and unit variance on a $500 \times 500$ pixel grid, approximately comparable to the galaxy data. Uncorrelated, isotropic Gaussian white noise is generated in Fourier space, smoothed using a Gaussian kernel with standard deviation $s$ (in pixel units), and then inverse-transformed. $s$ introduces a characteristic smoothing scale. Hereafter, we refer to this as Gaussian smoothing. Larger $s$ suppresses high-frequency (large $k$) modes, resulting in smoother, large-scale fluctuations. Conversely, a smaller $s$ retains high-frequency components, producing finer, small-scale structures. We vary $s$ to study how $H,~C$, and S$_{\textrm{gn}}$ change as functions of spatial scale. 

Figure~\ref{fig:gaussian} shows realizations of the GRF for a fixed random seed with varying $s$ (top row). We compute $H$ and $C$ for these GRFs, using embedding dimensions: $(d_x,d_y)=(2,2)$, (2,3), and (2,4). For all $(d_x,d_y)$ pairs, $H$ decreases with increasing $s$. 
This can be visually inferred from the fact that for low $s$ the GRFs contain more modes (hence more information and disorder), while for large $s$ the field becomes smoother and more ordered. For fixed $s$, $H$ have lower values for higher $(d_x,d_y)$, which is expected (see discussion in sec.~\ref{sec:s2a}). 

$C$ exhibits a rapid increase at low $s$, reaches a maximum (seen for the higher $(d_x,d_y)$ values in Figure~\ref{fig:gaussian}), and then tends to decrease. The curves for different embedding dimensions cross each other due to different rates of increase and subsequent decrease with increasing $s$. For improved visual clarity, we show these $H$ and $C$ results in semi-log plots in the left and middle panels of the second row of Figure~\ref{fig:gaussian}.

The right panel presents the $CH$-plane for the three embedding dimensions, with the black arrows indicating the direction of increasing $s$. 

{\em Limit of perfect correlation}: In the above discussion, we focus on smoothing scales $s$ that are in the ballpark range appropriate for the galaxy data in the next section. In the limit $s\to\infty$ (a completely
uniform field), the field tends to a constant function, and $H \to 0,\, C\to 0$.

We now map the GRFs at each $s$ to ordinal networks and calculate
the horizontal and vertical global node entropies, S$_{\rm gn}^{\rm hor}$ and S$_{\rm gn}^{\rm ver}$, respectively. In the bottom row of Figure~\ref{fig:gaussian}, we plot the variation of S$_{\rm gn}^{\rm hor}$ (dotted), and S$_{\rm gn}^{\rm ver}$ (dashed) with $s$ (left), along with their difference, $\Delta$S$_{\rm gn}$ (middle). $\Delta$S$_{\rm gn}$ decreases with $s$ and approaches zero, capturing directional asymmetries at small scales. Next, in the right panel, we fit an inverse power-law function (solid green line) to $\Delta$S$_{\rm gn}(s)$ (green triangles) in a log-log plot, yielding $\Delta$S$_{\rm gn}(s)$=18.58 $\times  s^{-1.18}$ with a goodness-of-fit value of R$^2=$0.166, indicating not particularly good fit. Investigating similar relations in the NGC 628 data may reveal signatures of self-organization arising from multiscale physical interactions in the galaxy. 

\section{NGC 628 Galaxy - multiwavelength view and observed data}
\label{sec:s3}

The galaxy NGC 628 is a nearly face-on (i=8.9$^\circ$; \cite{Lang_2020}) grand design spiral galaxy (SA(s)C; \cite{1991rc3..book.....D}) at a distance of 9.84 Mpc \citep{10.1093/mnras/staa3668}. It is a massive galaxy (log($M_{*}/M_{\odot}$)=10.34;  \cite{2019ApJS..244...24L}) with extended neutral hydrogen and UV disk~\citep{1992A&A...253..335K, 2007ApJS..173..538T, Yadav_2021}. 

In this section, we describe the multiwavelength imaging data used in our study and also overview physical features observed across the wavelengths of NUV, NIR, MIR, and mm, as shown in Figure~\ref{fig:fig2}(a). 

\subsection{Multiwavelength observed data of NGC 628}
\label{sec:sec3a}

We use archival UV observations from the Ultraviolet Imaging Telescope (UVIT) on board the AstroSat telescope~\citep{2012SPIE.8443E..1NK}. UVIT consists of two co-aligned Ritchey–Chrétien telescopes with a field of view of 0.5\textdegree, one for FUV (1300–1800 Å) and the other for NUV (2000–3000 Å) and visible (VIS) bands. We obtain Level 1 data from the Indian Space Science Data Centre (ISSDC) website~\footnote{\fontsize{6pt}{12pt}\selectfont \url{https://astrobrowse.issdc.gov.in/astro_archive/archive/Home.jsp}}. 
The raw UVIT data is reduced following~\cite{Amrutha_2025}. Multiple filter observations are available for NGC 628. We use only the NUVB4 N263M filter (1365s exposure time).

NGC 628 is a part of the Physics at High Angular Resolution in Nearby GalaxieS (PHANGS) program~\footnote{\fontsize{6pt}{12pt}\selectfont \url{https://sites.google.com/view/phangs/home}}. 
We use reduced archival data from the Atacama Large Millimeter/submillimeter Array (ALMA;~\cite{Leroy_2021a, Leroy_2021b}) CO ($J=2\to 1$) moment 0 map and the James Webb Space Telescope (JWST;~\cite{Lee_2023}) MIR instrument (MIRI) F2100W~\footnote{\fontsize{6pt}{12pt}\selectfont \url{https://jwst-docs.stsci.edu/jwst-mid-infrared-instrument/miri-observing-modes/miri-imaging\#gsc.tab=0}} and NIR camera (NIRCam) F360M images, both obtained from the PHANGS Treasury. Table~\ref{table1} lists relevant image parameters for the data used in our study.

\begin{table}[h!]
\centering
\small 
\caption{Image parameters}
\begin{tabular}{lcccc}
\hline
\ \ Image & Central & Image  &Platescale & Pixel unit\\
& wavelength& PSF & ("/pix) & \\
  &  (nm) &(")  & & \\
\hline
UVIT NUV  &$2.63\times 10^{2}$& 1.2 & 0.417&counts\\
JWST NIR &$3.6\times 10^{3}$    &0.12 & 0.063&MJysr$^{-1}$\\
JWST MIR &$2.1\times 10^{4}$ &0.67 & 0.11&MJysr$^{-1}$\\
ALMA  mm 
&$1.3\times 10^{6}$& 1& 0.2& Kkms$^{-1}$\\
\hline
\end{tabular}
\label{table1}
\end{table}
\normalsize

\begin{figure*}
\centering
(a)\\
\includegraphics[scale=0.67]{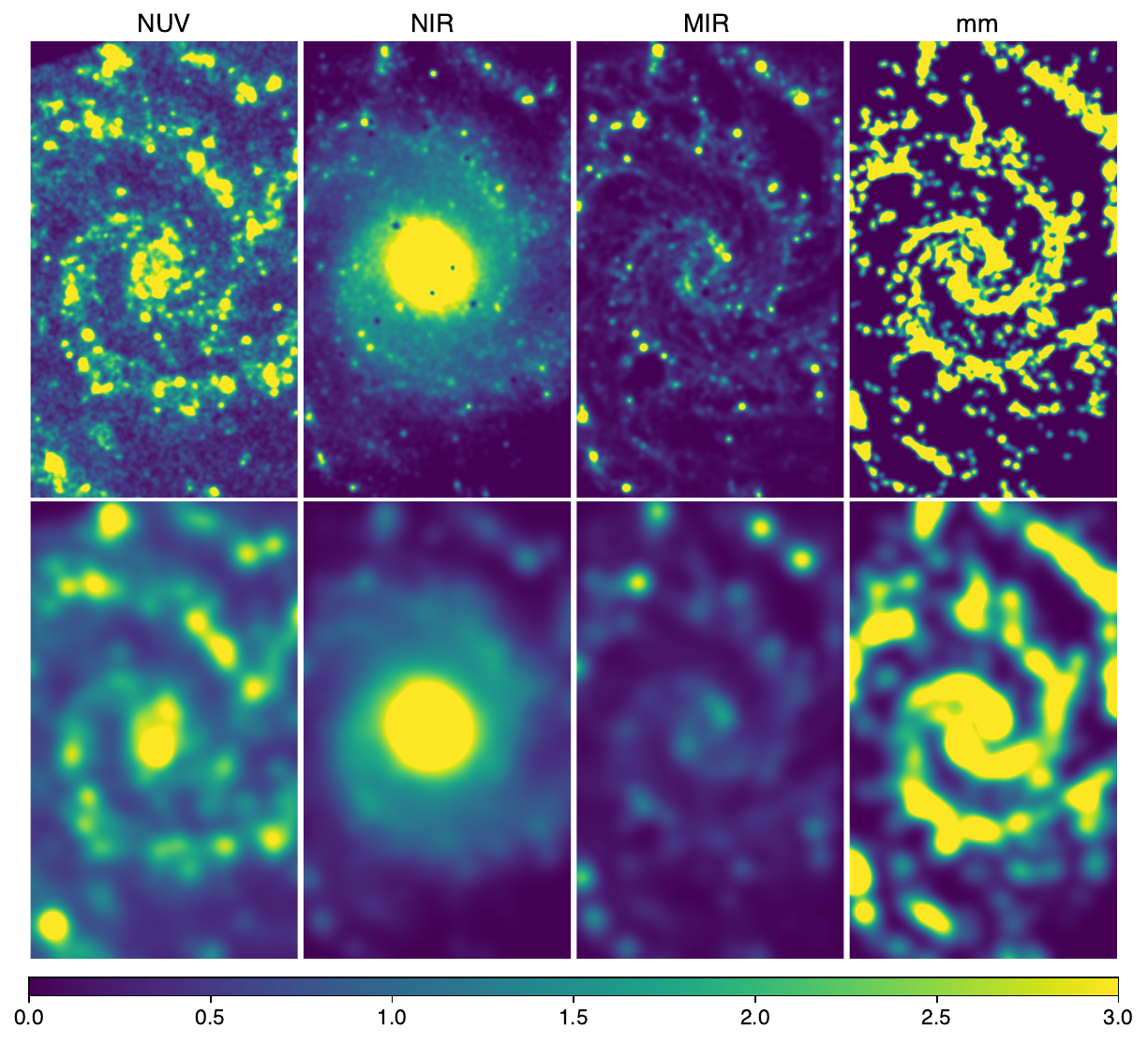}\\
(b)\\
\includegraphics[scale=0.24]{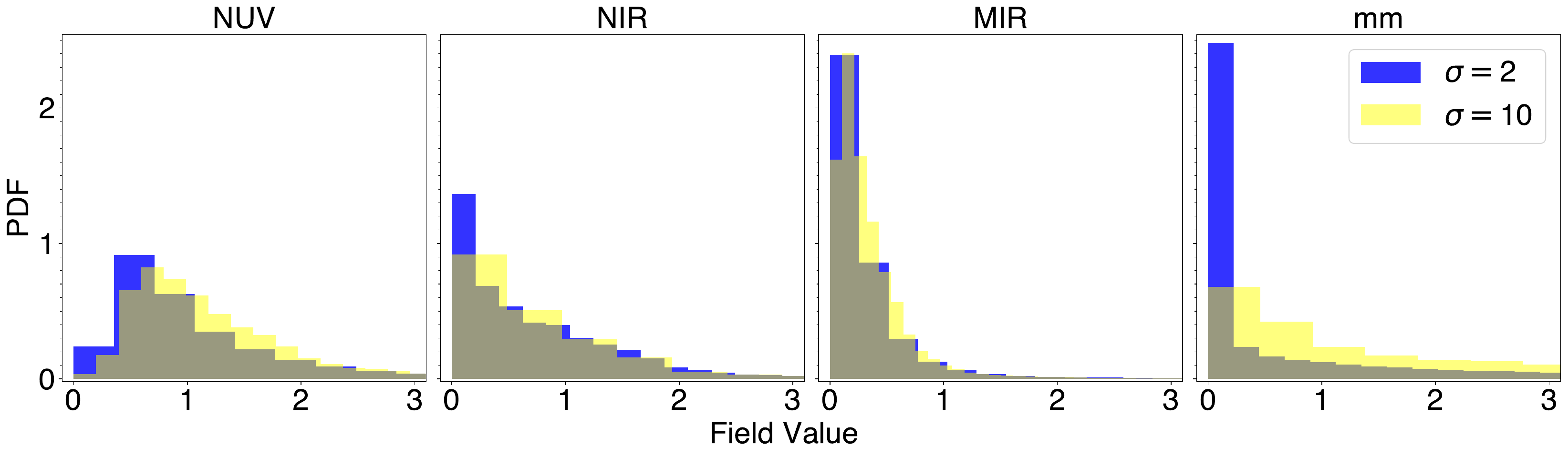}
\caption{{\em Top row of (a)}: Images of the galaxy NGC 628 observed at NUV, NIR, MIR, and mm wavelengths (left to right) obtained from UVIT, JWST (NIR and MIR), and ALMA, respectively. All images cover identical fields of view. The intensity values of each image have been normalized by the respective standard deviation. All images are smoothed identically using a Gaussian kernel with a smoothing scale $\sigma=2$ (in pixels). {\em Bottom row of (a)}: Same as the top row but for $\sigma=10$. Image parameters are provided in  table~\ref{table1}. (b) PDFs of the field values of the four images in (a) corresponding to the top row (blue), and bottom row (yellow).} 
\label{fig:fig2}
\end{figure*}

To ensure consistent spatial scales across all four images, we apply interpolation so that the plate size would cover the same area of the galaxy ($0.417^{"}/{\rm pix}$). 
Each pixel corresponds to a physical size of 19.89 pc. We convolve all images to a common resolution of $1.2^{''}$ (UVIT) (the poorest resolution among the maps), corresponding to the physical scale of 57 pc. The final images, displayed in Figure~\ref{fig:fig2}(a), consist of $530 \times 311$ pixels, spanning $10.54 \times 6.19\, {\rm kpc}^2$.

\subsection{Multiwavelength view of NGC 628}
\label{sec:3b}

\textit{NUV}: This emission traces recent star formation, predominantly from O, B, and massive A stars over timescales of 200-300 Myr \citep{2005ApJ...619L..79T}. In NGC 628, large NUV star-forming complexes are found in the wider arms and inter-arm regions. A faint UV emission is observed across the disk, especially at the galaxy's center, indicating some evolved stellar populations and scattered dust. NGC 628 has experienced self-regulated star formation over the last 500 Myr~\citep{1994ApJ...426..553C, 10.1093/mnrasl/slz093}, owing to the lack of interactions in the past Gyr.

{\em NIR} (3.6 $\mu$m): This emission traces the bulk of low-mass stars and evolved giants (which have moved off the main sequence), producing smoother, less clumpy arms. The center is bright, with two prominent spiral arms and a visible bifurcation in the top-right arm.

\setcounter{footnote}{0}

{\em MIR:} The emission in MIR is due to re-emission of UV and visible light absorbed by the dust, tracing embedded star formation. The MIRI F2100W band (includes 24 $\mu$m reveals bright spiral arms, spurs \citep{Williams_2022}, bubbles \citep{Watkins_2023}, and web-like inter-arm structures.

{\em Millimeter (mm)}: The mm emission of our data corresponds to the $J=2\to 1$ transition of carbon monoxide (CO) molecules, which indirectly traces cold molecular hydrogen clouds that exhibit clumpy structures. With the ALMA strict map, diffuse emission is filtered out~\citep{Leroy_2021a}. CO emission is bright throughout the spiral arms and spurs.

\subsection{Coarse graining images }

To carry out a multiscale probe of the structural properties of NGC 628, we systematically perform coarse-graining (smoothening) of the multi-wavelength images using Gaussian kernels with different standard deviations $\sigma$ (in units of pixel numbers). Throughout this paper, we use `smoothing scale' to mean the standard deviation of the smoothing kernel. This process mimics telescope beam smoothing and enables analysis at varying resolutions. By varying $\sigma$, we study small and large-scale morphological features arising from multiscale physical interactions. Figure~\ref{fig:fig2}(a) displays the images smoothed with $\sigma=2$ (first row) and 10 (second row). 

{\em Effective physical scale}: The point spread function (PSF) values quoted in table~\ref{table1} are full width at half maximum (FWHM) values. The common smoothing of all images by FWHM 1.2" corresponds to 57 pc in physical scale, and the corresponding smoothing scale is 24.2 pc. In pixel units, this corresponds to a smoothing scale, denoted by $\sigma_a$, of 1.22.  Since the smoothing by $\sigma$ discussed above is in addition to the common smoothing by $\sigma_a$, the effective smoothing scale, in pixel units,  is
\bea
\sigma_{\rm eff} &=& \sigma\sqrt{1+(\sigma_a/\sigma)^2}.
\eea
For $\sigma\gg\sigma_a$, we have $\sigma_{\rm eff}\sim\sigma$. 
Then the physical length scales, which we denote by $L_{\rm FWHM}$, associated with the FWHM value corresponding to $\sigma_{\rm eff}$ is given by
\be
L_{\rm FWHM}=2\sqrt{2\ln 2}\times\sigma_{\rm eff}\times 19.89\,{\rm pc}.
\label{eq:Reff}
\ee
In section~\ref{sec:s4} we will primarily use the smoothing scale $\sigma$ for plotting and presenting our results, unless stated otherwise. However, for interpretation in terms of physical length scales, we will use  $L_{\rm FWHM}$, which is the length corresponding to the smoothing FWHM.  

\subsection{Probability distribution functions at multiwavelengths}

The panels in Figure~\ref{fig:fig2} (b) show the probability density functions (PDFs) of the emissions of NGC 628 at NUV, NIR, MIR, and mm wavelengths, for smoothing scales $\sigma=2$  (blue) and $\sigma=10$ (yellow). We observe that all PDFs are right-skewed. For the mm image, we observe a sharp drop in the PDF values for lower field values. A precise modeling of the PDFs is not important for our discussion here. 

\section{Results for NGC 628}
\label{sec:s4}

In this section, we present the results obtained for permutation entropy $H$, disequilibrium $D_E$, statistical complexity $C$, and ordinal network global node entropy S$_{\rm{gn}}$ for the multiwavelength images of NGC 628. We probe the multiscale nature by smoothing the galaxy field data at different scales $\sigma$ and using different choices of embedding dimensions $(d_x,d_y)$.

{\em Choice of embedding dimensions}: We use combinations of $(d_x,d_y)=$ (2,2), (2,3), and (2,4). These choices are constrained by the total number of available pixels in the data, and the condition for statistical reliability: $(d_xd_y)!\ll 164,830$. Different embedding dimensions explore correlations for varying chunks of data, and this can potentially reveal distinct structural features that may not be apparent at a single embedding dimension. 

\begin{figure}
\centering
\includegraphics[scale=0.25]{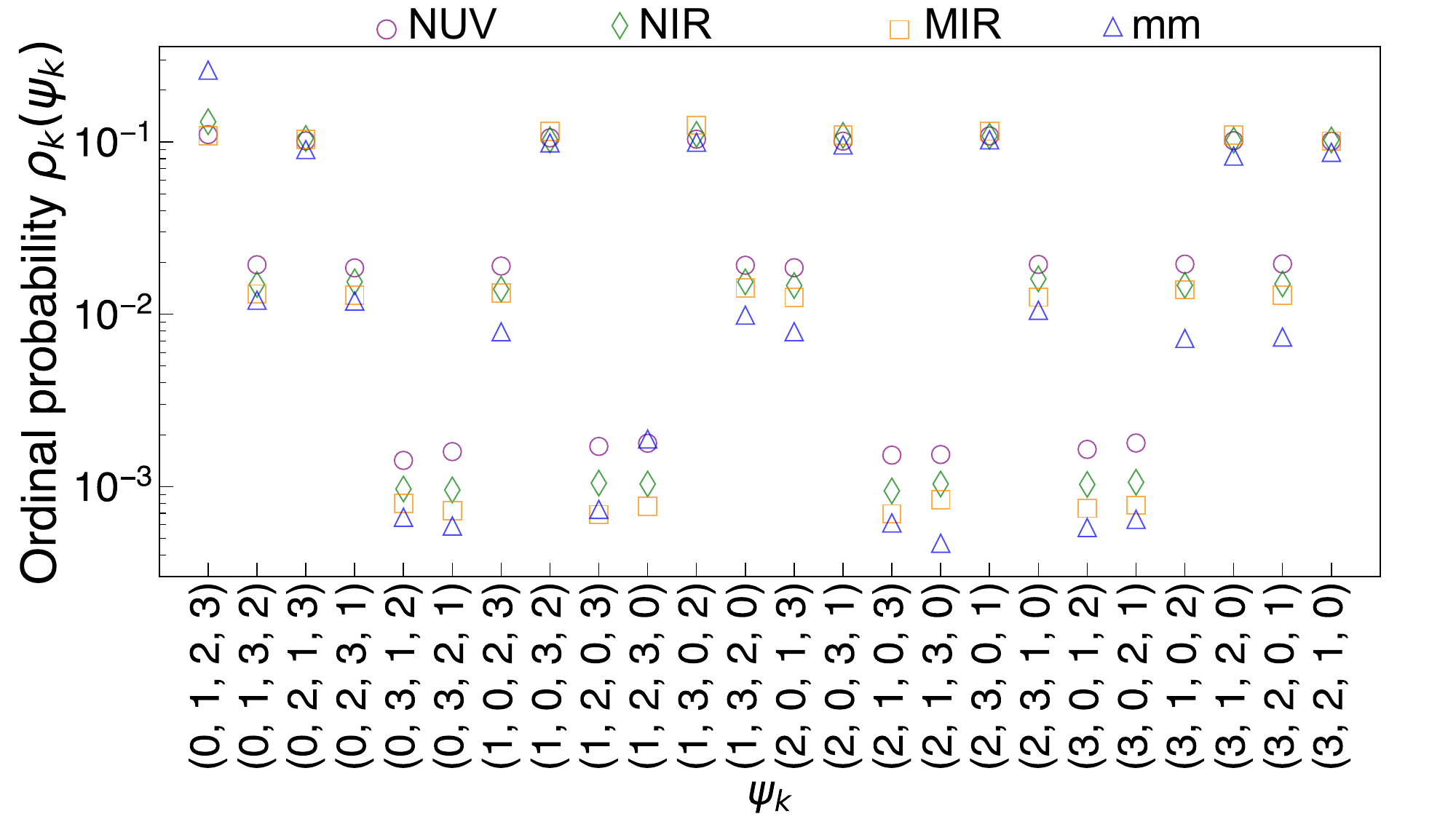}
\caption{Ordinal probabilities $\{\rho_k(\psi_k)\}$ of the four NGC 628 images  when $(d_x,d_y)=(2,2)$ at smoothing scale $\sigma=2$.}
\label{fig:P}
\end{figure}

The numbers of distinct states, permutation states, and the physical sizes (each pixel rounded off to 20 pc)  associated with the three $(d_x,d_y)$ choices are listed as: 
\vskip .1cm
\noindent
\begin{tabular}{cccc}
\hline
$(d_x,d_y)$ &  \texttt{No. of  } & \texttt{No. of }  &  \texttt{Physical }  \\
&\texttt{distinct}&\texttt{permutation}&\texttt{size in}\\
&\texttt{states}&\texttt{states}&${\rm pc}^2$\\
\hline
(2, 2) & 24   &  163,990  & $40\times 40$  \\
(2, 3) &  720   &  163,461  & $40\times 60$ \\
(2, 4) & 40,320   & 162,932 & $40\times 80$  \\
\hline
\end{tabular}
\vskip .3cm
As a demonstration, we present the ordinal probabilities $\{\rho_k(\psi_k)\}$
for the four images of NGC 628 at $\sigma=2$ for $(d_x,d_y)=(2,2)$ in  Figure~\ref{fig:P}.  The $x$-axis shows the 24 distinct states $\psi_k$. We see that some states are more probable than others. The manner in which the ordinal probabilities deviate from equi-probable distribution results in distinctive behavior of the statistical quantities ($H$, $D_E$, $C$, and S$_{\rm{gn}}$) introduced in section~\ref{sec:s2}. 

{\em Estimation of errors}: We estimate errors associated with the computed values of $H,\,D_E$ and $C$ values of NGC 628 as follows. We divide the data into 20 equal-sized patches, each having an area roughly $2.11\times 1.53$\,${\rm kpc}^2$. Then, treating each patch as independent data, we compute $H,\,D_E$ and $C$, and treat the standard deviations from the 20 values as the statistical uncertainties. Note that this way of estimating errors will slightly underestimate the true errors since the different patches are actually correlated.

The errors in $H$ and $D_E$ will propagate to yield the error in $C$.  To quantify this, let $\sigma_H,\,\sigma_{D_E}$ and $\sigma_C$ denote the standard deviations (errors). Then, 
\be
\sigma_C^2= \bar D_E\sigma_{H}^2 +\bar H\sigma_{D_E}^2 +2\bar H\bar D_E {\rm Cov}\left(H,D_E\right),
\ee
where the bar indicates mean values, and ${\rm Cov}\left(H,D_E\right)$ is the covariance between $H$ and $D_E$. Intuitively, we expect $H$ and $D_E$ to be anti-correlated, with ${\rm Cov}\left(H,D_E\right)$ being negative. We explicitly compute the covariance and confirm that this is so. Then, $\sigma_C$ will not be simply related to the quadrature sum of   $\sigma_{H}$ and $\sigma_{D_E}$.  
For all results shown henceforth we will use $\sigma_H,\,\sigma_{D_E}$ and $\sigma_C$ to show the error bars.

\subsection{Entropy, Disequilibrium, and Complexity 
for different wavelength images of NGG 628 across scales} 
\label{sec:4a}

\begin{figure*}
\centering
\includegraphics[scale=0.7]
{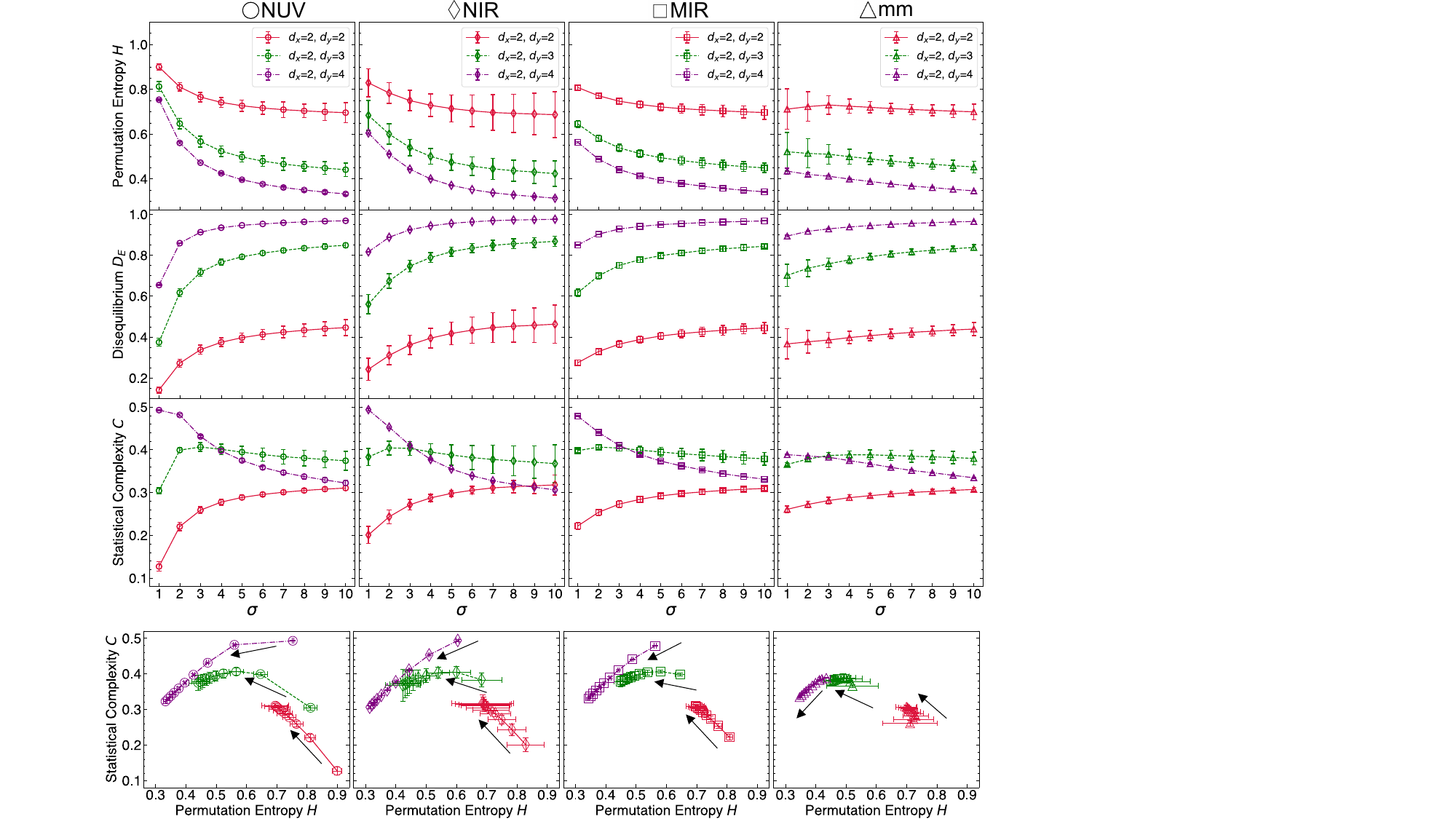}
\caption{{\em First, second and third rows}: $H, D_E$ and $C$ computed using $(d_x,d_y)=$ (2,2) (red), (2,3) green and (2,4) (purple) for the four images of NGC 628 shown in Figure~\ref{fig:fig2}  smoothed at different scales $\sigma$ (in units of 10\,pc). The details of the images are provided in  table~\ref{table1}. {\em Bottom row}: Values of $C$ and $H$ shown on the $CH$-causality plane for the three cases of $(d_x,d_y)$ of the four NGC 628 images at different smoothing scales. The black arrows indicate the direction of increasing $\sigma$. } 
\label{fig:fig3}
\end{figure*}

Figure~\ref{fig:fig3} shows permutation entropy ($H$), disequilibrium ($D_E$), and statistical complexity ($C$) as functions of smoothing scale $\sigma$ at $(d_x,d_y)=(2,2),~(2,3)$ and $(2,4)$. From the top row, at all embedding dimensions $(d_x,d_y)$, in all bands except mm, $H$ decreases with increasing $\sigma$ consistent with the loss of small-scale structures seen in the images: in NUV emission from evolved stellar population and dust scattering, in NIR due to starlight from low-mass stars, dust continuum, and scattered light, and in MIR from bright clumpy dust in the spiral arm, with diffuse interarm web structures. Since $H$ quantifies randomness, diffuse structures in the observed data have a pronounced effect on $H$. At mm wavelength, however, $H$ shows only mild variation with $\sigma$, reflecting the cold and clumpy nature of the emission and the lack of diffuse structures, as also evident in the PDFs of Figure~\ref{fig:fig2}(b), where mm and MIR have most of the field values around 0. A mild non-monotonic behavior for $(d_x,d_y)=(2,2)$ at small $\sigma$ is further explained by the clumpy nature of regions that emit in mm wavelengths, making them less disordered. The second row shows increasing $D_E$ with $\sigma$ across all wavelengths and $(d_x,d_y)$, indicating greater structural order as fluctuations are washed out with smoothing. The third row shows the behaviour of $C$, encoding information of both $H$ and $D_E$, with $\sigma$ at different $(d_x,d_y)$, highlighting that different embedding dimensions capture distinct aspects of the galaxy's complexity inherent in different spatial structures of the galaxy field at the given scale.

We observe that for $H$ and $D_E$, $(d_x,d_y)=(2,2)$ has the largest error bars, with a drop towards higher embedding dimensions across all wavelengths. The errors tend to increase with $\sigma$, except for mm, which shows the opposite trend. Moreover, NIR shows larger error bars relative to other wavelengths, indicating larger fluctuations across the galaxy. MIR shows the smallest error bars, indicating relative uniformity of the fluctuations across the galaxy. 
We also observe that for $C$ the sizes of error bars do not follow the trends of $H$ and ${D_E}$, which is because of  ${\rm Cov}\left(H,D_E\right)$ being negative. 

The bottom row of Figure~\ref{fig:fig3} shows the $CH$-planes parametrised by $\sigma$, with the black arrows indicating increasing $\sigma$. Across different embedding dimensions,  $H$ consistently decreases with increasing smoothing for all images (except mm at $(d_x,d_y)=(2,2)$). However, complexity $C$ exhibits interesting behaviors: for $(d_x,d_y)=(2,2)$, $C$ shows increasing trends; for $(d_x,d_y)=(2,4)$, $C$ decreases; and for $(d_x,d_y)=(2,3)$, $C$ follows non-monotonic behaviors. Since $C$ ($=D_E\times H$) reflects the interplay between order (organization) and disorder (randomness), these variations of $C$ reflect distinct complexity inherent in different spatial structures and scales of the galaxy. These results, therefore, highlight the morphological complexity of NGC 628 at multiscales.

\subsection{Entropy, Disequilibrium, and Complexity of NGC 628 across observing wavelengths}
\label{sec:s5b}

\begin{figure}[h]
\includegraphics[scale=0.3]
{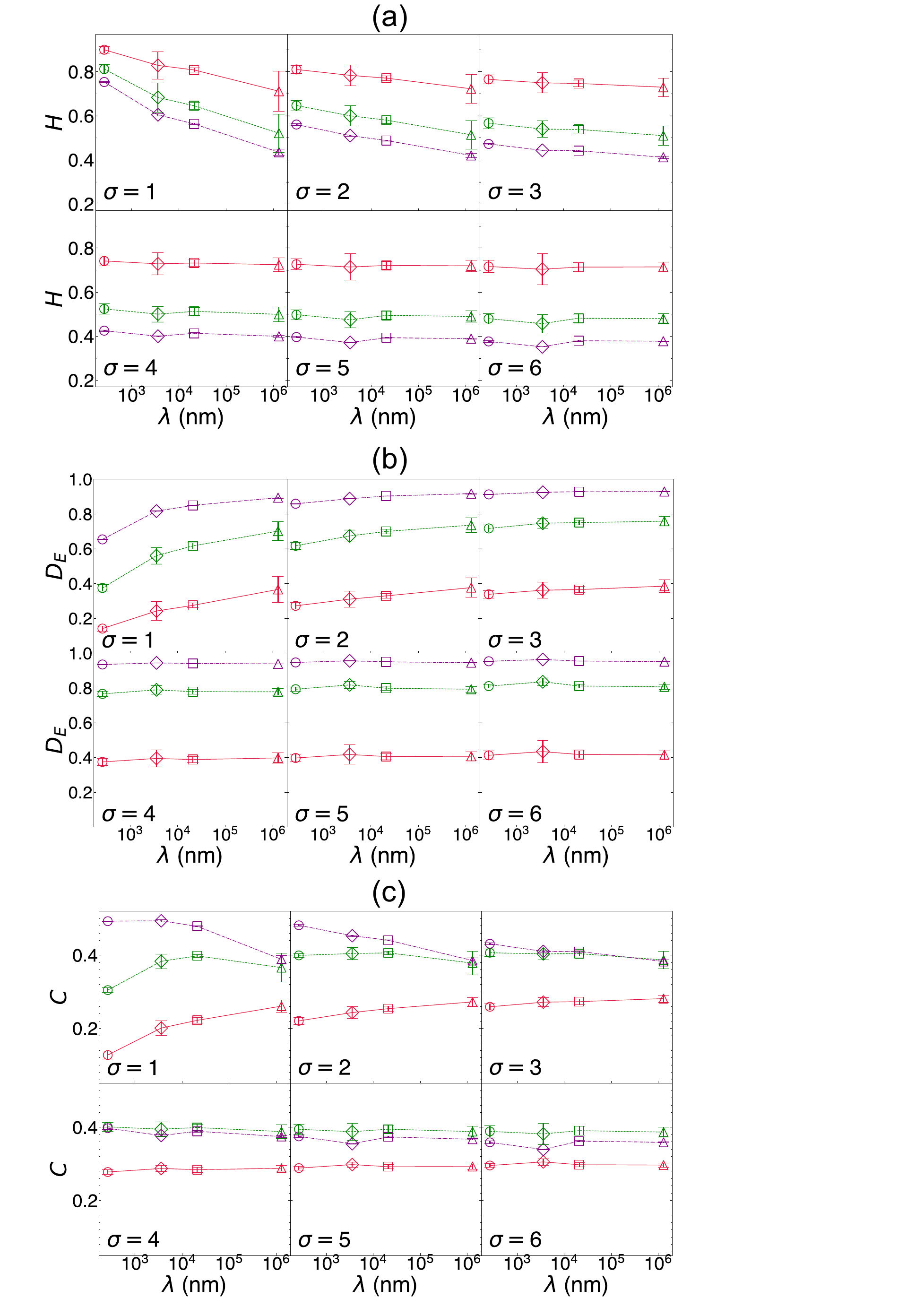}
\caption{Comparison of $H$ ((a)), $D_E$ ((b)) and $C$ ((c)) as functions of observing wavelength $\lambda$ (in nm), for each smoothing scale.}
\label{fig:fig4}
\end{figure}

\begin{figure}
\centering
\includegraphics[scale=0.5]{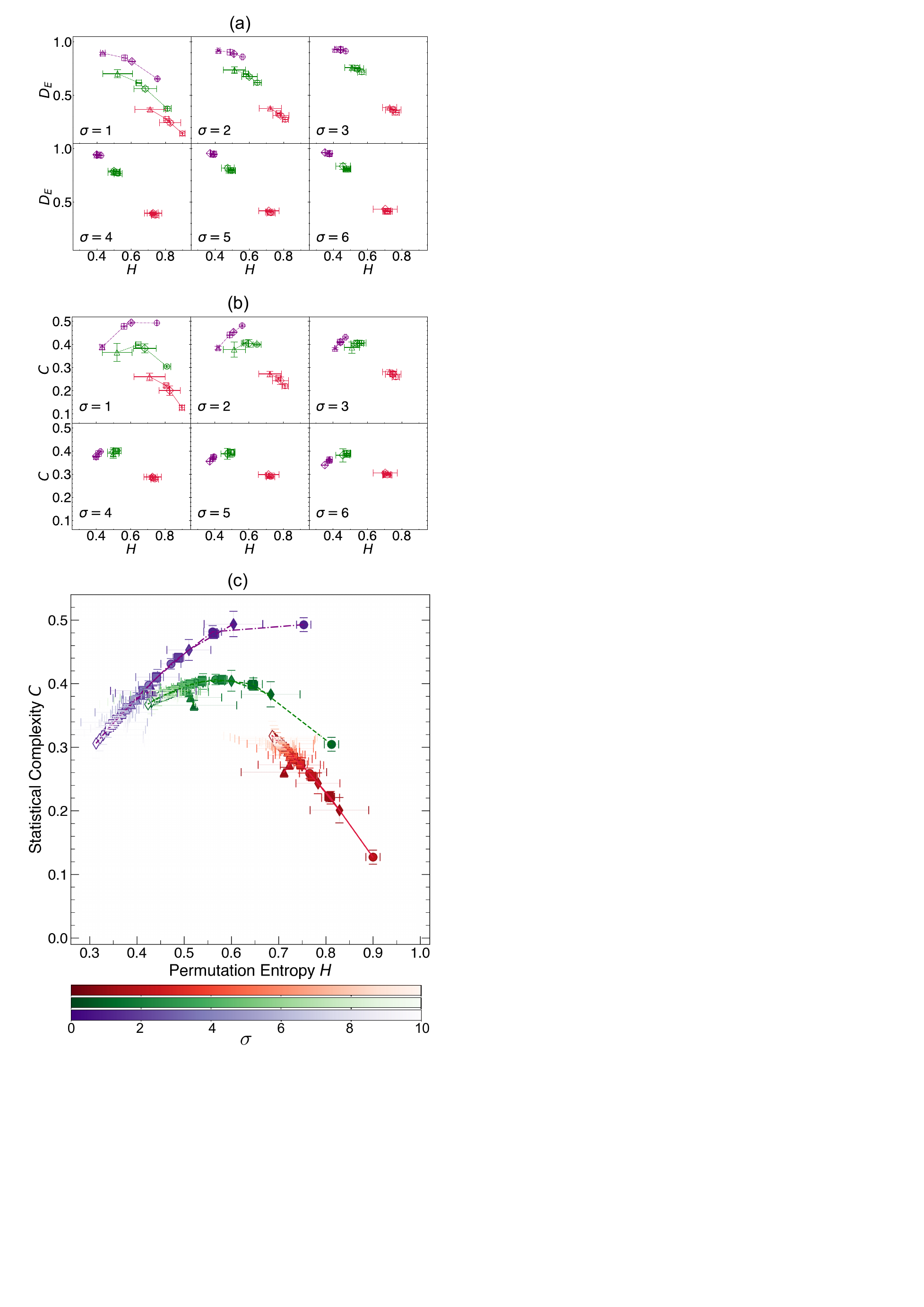}
\vspace{-4.5cm}
\caption{(a) $D_E$ versus $H$ for the four images of NGC 628  shown in the same panel (same markers as Figure~\ref{fig:fig4}), for different $\sigma$ and ($d_x, d_y$). (b) $CH$-plane for different $\sigma$. 
(c) $C$ and $H$ values for all wavelengths and $\sigma$ are plotted on the same $CH$-plane. The color bars at the bottom indicate larger $\sigma$ towards lighter hues.
}
\label{fig:fig5}
\end{figure}

\begin{figure*}
\centering
\includegraphics[scale=0.41]{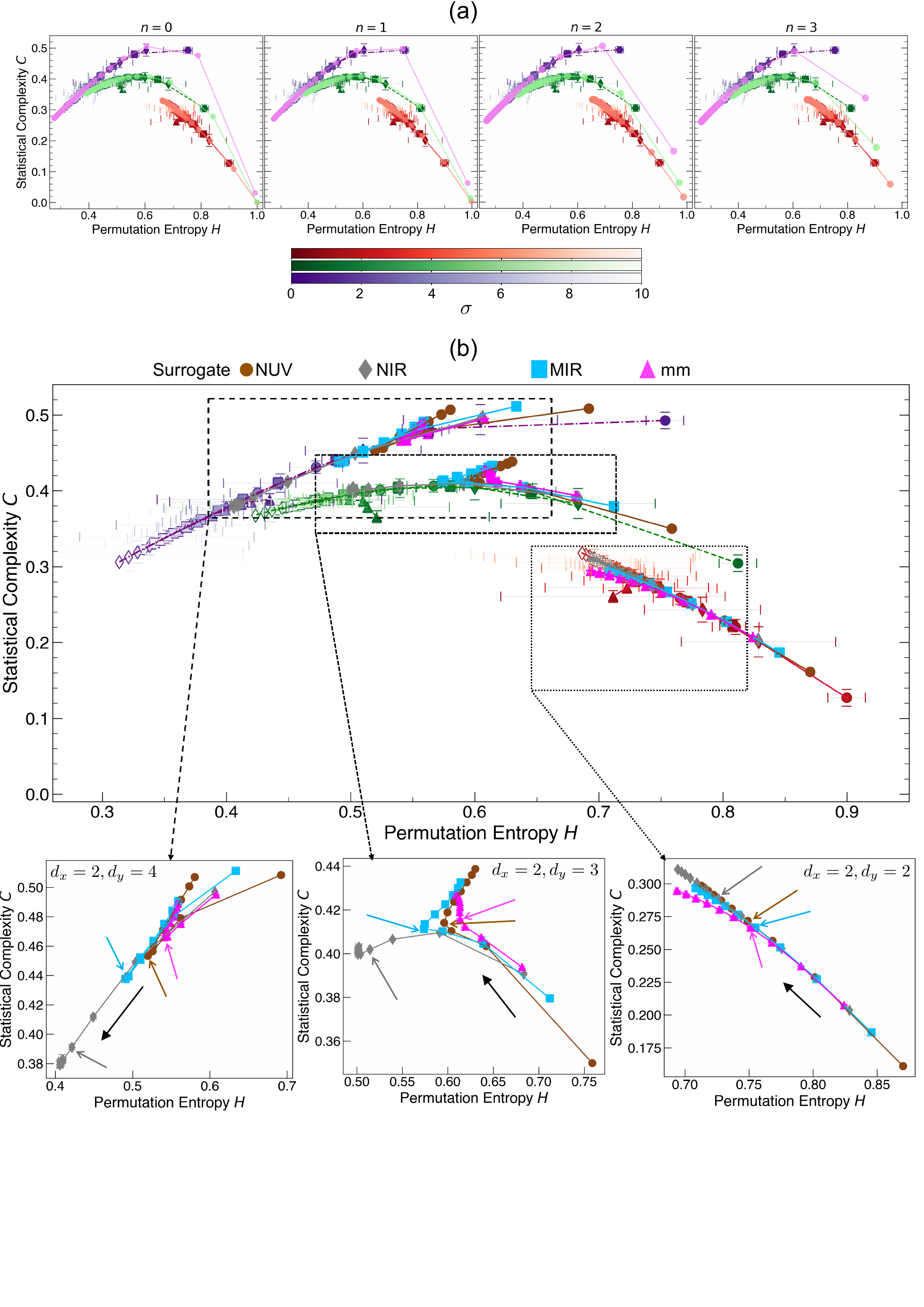}
\vspace{-3cm}
\caption{(a) $CH$-planes for comparison between the four wavelength images of NGC 628 and GRFs with power spectra of the form $P(k)\propto k^{-n}$ with $n=0,1,2,$ and 3. In each panel, we use the same markers and colors (indicated by the color bar) as in Figure~\ref{fig:fig5}(c) for the four wavelengths, whereas the lighter-colored curves now correspond to the GRFs at the three different embedding dimensions $(d_x,d_y)=(2,2), (2,3)$ and $(2,4)$. (b) \textit{Surrogate fields:} In the upper panel, we plot the $(C,H)$ values of the four images of NGC 628 along with the corresponding plots for the surrogate fields (indicated by different markers and colors at the top of the panel). Each surrogate field is a GRF constructed by preserving the power spectrum of the galaxy and randomizing the Fourier phases, at each wavelength and smoothing scale $\sigma$. The bottom panels show the zoomed-in views of the above boxed regions, showing the $(C,H)$ values of the surrogate fields only (for visual clarity) at the three different embedding dimensions $(d_x,d_y)=(2,2), (2,3)$ and $(2,4)$. In the panels, each point on the $CH$-plane corresponds to $\sigma$ values from 1 to 10 (increasing in the direction indicated by the bold black arrows) for a given surrogate field (same marker and color). Coloured arrows (brown, grey, light blue, and pink) indicate where $\sigma=4$ is located.}
\label{fig:surr}
\end{figure*}

We next examine how $H,~D_E$, and $C$ vary with observing wavelength $\lambda$. Figure~\ref{fig:fig4} shows $H,~D_E$, and $C$ versus $\lambda$ for $\sigma=1,2,\dots,6$. The markers and colour codes are the same as in Figure~\ref{fig:fig3}. 
Figure~\ref{fig:fig4} reveals two different behaviors in the variation of $H,~D_E$, and $C$ with $\sigma$ for all $\lambda$. For scales $\sigma<4$, $H$ decreases with increasing wavelength for all embedding dimensions, as seen in panel (a). NUV has the highest entropy, which is explained by the fluctuating nature of star-forming regions across a wide range of spatial scales with diffuse emission in the entire disk. In comparison, there is a drop in diffuse emissions in NIR and further in  MIR, resulting in lower values of $H$. The mm image has the lowest $H$ value because the emissions primarily originate from cold, clumpy regions, with no diffuse emissions. Besides, UV emission from massive star associations can photoionize the surrounding gas \citep{1996ApJ...460..914V}, making fluctuations particularly prominent in the NUV band. Previous studies have also indicated that these associations drive large-scale motions, contributing to turbulence and structure in the interstellar medium (ISM)~\citep{2004ARA&A..42..211E, kim2007topology}. Through stellar winds and supernova explosions, these stars inject energy~\citep{1994ARA&A..32..227M}, and create bubbles and superbubbles, ranging from a few parsecs to several kiloparsecs, which are evident in the JWST MIR observations of NGC 628~\citep{Barnes_2023, Watkins_2023}. However, for scales $\sigma\ge 4$, $H$ does not significantly vary with the variation of wavelength as both small-scale structures and diffuse emissions get washed out due to the effect of larger smoothing. Further examination of $D_E$ and $C$ also shows similar constant behavior as $H$ for $\sigma\ge 4$.

{\em Emergence of a transition scale}: 
In Figure~\ref{fig:fig5}(a) and (b), we present the $D_{E}H$- and $CH$-planes, where data points corresponding to all four wavelengths (distinguished by different markers) are plotted together for each value of $\sigma$. We observe that, for all embedding dimensions, the values across the four wavelengths converge for $\sigma \ge 4$. This observation reinforces the conclusion drawn earlier from Figure~\ref{fig:fig4}. Hence, based on visual inspection, $\sigma = \sigma^{(c)} \sim 4$ emerges as a critical transition scale beyond which the degrees of order and disorder in NGC 628 exhibit universal behavior across all four wavelengths.  Converting $\sigma^{(c)}$ to a physical scale, which we denote by $L_{\rm FWHM}^{(c)}$ using eq.~\ref{eq:Reff}, we obtain 
a value of 196 pc as the effective transition length scale. We note that this value is not precisely determined, and the uncertainty has not been rigorously quantified. However, a rough estimate using $L_{\rm FWHM}(\sigma=3) \simeq 152$ pc gives the error to be $[(196-152)/2]$\,pc, which is 22\,pc.  Hence the uncertainty must be around ten percent.  
For this reason we quote the value of $L_{\rm FWHM}^{(c)}$ to be of order $\sim 200$ pc.

{\em Physical significance of the transition scale}: At scales below $L^{(c)}_{\rm FWHM}$, NUV has the highest $H$ and lowest $D_E$, with decrease/increase of $H/D_E$ towards higher wavelengths. This trend supports the physical expectation that UV photons generated by stars transfer energy to the ISM, and have a cascading (but delayed in terms of a shift in the values of $H$ and $D_E$) effect. There exists a hierarchy of scales related to star clusters such as OB associations and stellar aggregates (see e.g.~\cite{Gusev:2014}); and characteristic scales associated with distinct features of the ISM such as clumps and holes related to stellar winds, supernova bubbles and super-bubbles (arising from multiple supernova explosions)~\citep{kim2007topology}. The hierarchy of these scales is revealed as the increasing/decreasing part of the plots of $H/D_E$ versus $\sigma$. At $L^{(c)}_{\rm FWHM}$ and higher, all wavelengths share the same values of $H$ and $D_E$, implying that they trace the same structural features. Therefore,  $L^{(c)}_{\rm FWHM}\sim 200$\,pc captures the transition from scales affected by star formation and stellar feedback, to larger scales where the galaxy's morphology is shaped by global gravitational dynamics.  

{\em Emergence of attractor curves}: Figure~\ref{fig:fig5}(c) shows the values of $C$ and $H$ for all wavelengths (distinguished by markers) across smoothing scales, plotted on a common $CH$-plane. The progression toward lighter colors indicates increasing smoothing scales, as shown by the color bars at the bottom. We see that $C$ and $H$ values tend to align along the same curve at each $(d_x,d_y)$, except the mm image towards small $\sigma$. This suggests the existence of an {\em attractor curve} for NGC 628, for each ($d_x,d_y$). It is noteworthy that for the smallest embedding dimension $H$ and $C$ 
span relatively smaller regions of the phase space, indicating that larger embedding dimensions are necessary to better probe the structural components of the galaxy. 

Further, we generate GRFs with power spectra of the form $P(k)\propto k^{-n}$, where $n=0,1,2,$ and 3. The three light-colored $CH$-curves on each panel of Figure~\ref{fig:surr}(a) represent the $(H,C)$ values of smoothed GRFs (parametrised by $s$ and increasing in the same direction as $\sigma$) for fixed seeds corresponding to the three embedding dimensions. In each panel, we use the same markers and colors (indicated by the color bar) as in Figure~\ref{fig:fig5}(c) for the four wavelength images. We do not make a comparison between $s$ and $\sigma$ since the GRF is not a physical field, and we cannot compare the pixel sizes directly. What is of interest to observe is that the $CH$-trajectories of the GRFs with $n=1$ align most closely with those of the NGC 628 galaxy field, indicating that the galaxy field shares similar correlation properties with a pink-noise ($1/f$) spectrum. It further suggests an underlying universality of the attractor curves, despite the differing physical processes traced by each wavelength.

{\em Robustness of our results}: To test the effect of rotation of the data, we repeat all calculations for the four image data, as well as the GRFs, after interchanging $d_x$ and $d_y$ (except (2,2), which is symmetric). We find a negligible effect of the rotations on all the results. We note that the results would remain invariant to any order-preserving transformation of the data, since the ordinal distributions are the same.

\subsection{Surrogate analysis as a control study to quantitatively determine the transition scale}
\label{sec:surr}
Next, to test whether the galaxy’s $CH$-curve is determined solely by its power spectrum, we construct realizations of phase-randomized surrogate Gaussian random fields for each wavelength and smoothing scale $\sigma$. These surrogates preserve the exact power spectrum of the galaxy but have randomized Fourier phases, thereby destroying nonlinear correlations while retaining all linear correlations. Further justification on constructing these phase-randomized surrogates is provided in Appendix~\ref{sec:appendb}. In Figure~\ref{fig:surr}(b), we show the mean $(H,C)$ results (indicated by markers and colors at the top of the panel) obtained from 60 realizations of surrogate fields (with error bars) along with the galaxy results. The three panels in the bottom row show the zoomed-in views of the boxed regions. Black arrows indicate the direction of increasing $\sigma$. For $(d_x,d_y)=(2,2)$ (right panel), the phase-randomized surrogates track the galaxy data reasonably well, indicating that low-dimensional embeddings are dominated by linear correlations, which the power spectrum captures. However, for $(d_x,d_y)=(2,3)$ (middle panel) and $(2,4)$ (left panel), the $(H,C)$ values of the surrogate fields begin deviating from the galaxy curve already at small $\sigma$, and the deviation becomes particularly strong around $\sigma \sim 4$. This further demonstrates that $\sigma=\sigma^{(c)}\sim 4$ marks a critical transition scale, where nonlinear, phase-dependent correlations present in the galaxy data are no longer reproduced by Gaussian surrogates that match the power spectrum. We interpret the results as follows. Below $\sigma<\sigma^{(c)}$, surrogates lie close to the galaxy curve, and much of the small-scale structure is consistent with what the power spectrum captures for small ($d_x,d_y$). Nonlinear correlations (turbulence, fractal hierarchy) exist to cause significant deviations for larger $(d_x,d_y)$. Around $\sigma=\sigma^{(c)}$, surrogates begin to deviate significantly, marking the emergence of nonlinear, phase-dependent correlations that cannot be captured by the power spectrum alone. For $\sigma>\sigma^{(c)}$, deviations grow substantially. These scales reflect coherent galactic morphology, requiring a phase-organized structure that Gaussian surrogates cannot reproduce. We note that for NIR, $\sigma^{(c)}\sim 5$ and its $CH$-curve appears different as compared to the other wavelengths that trace younger stellar populations and molecular gas. Older populations are, in general, more dispersed and form larger complexes~\citep{2006ApJ...644..879E}. An additional bright, large structure at the center for NIR might be the reason behind the observed difference.

The transition scale $\sigma^{(c)}\sim 4$ is consistent with the correlation length, $l_{\rm corr}$, inferred using the two-point correlation function for NGC 628~\citep{Grasha_2017, 10.1093/mnras/stab2413}.  These studies interpret $l_{\rm corr}$ as the maximum separation over which star clusters remain correlated in the scale-free, turbulent fractal hierarchy generated by star formation and stellar feedback. Beyond this scale, the fractal hierarchy breaks down, and large-scale gravitational dynamics begin to shape global morphology. For a spiral galaxy like NGC 628, $l_{\rm corr}$ marks the shift from a fractal distribution influenced by turbulence to larger scales, where galactic dynamics dominate. Thus, below  $\sim$200 pc, the galaxy contains strong turbulent, fractal, nonlinear correlations, whereas above  $\sim$200 pc, the structure transitions to large-scale, phase-organized morphology that cannot be reproduced by Gaussian fields. The close agreement between $L^{(c)}_{\rm FWHM}$ and $l_{\rm corr}$ supports the interpretation that  $\sim$200 pc represents a fundamental physical transition scale in NGC 628: the boundary between turbulence-driven small-scale structure and large-scale galactic organization.

\subsection{Disorder to order transition for individual wavelength images of NGC 628  and high resolution JWST NIR and MIR images }
\label{sec:a3}

\begin{figure*}
\centering
\includegraphics[scale=0.43]{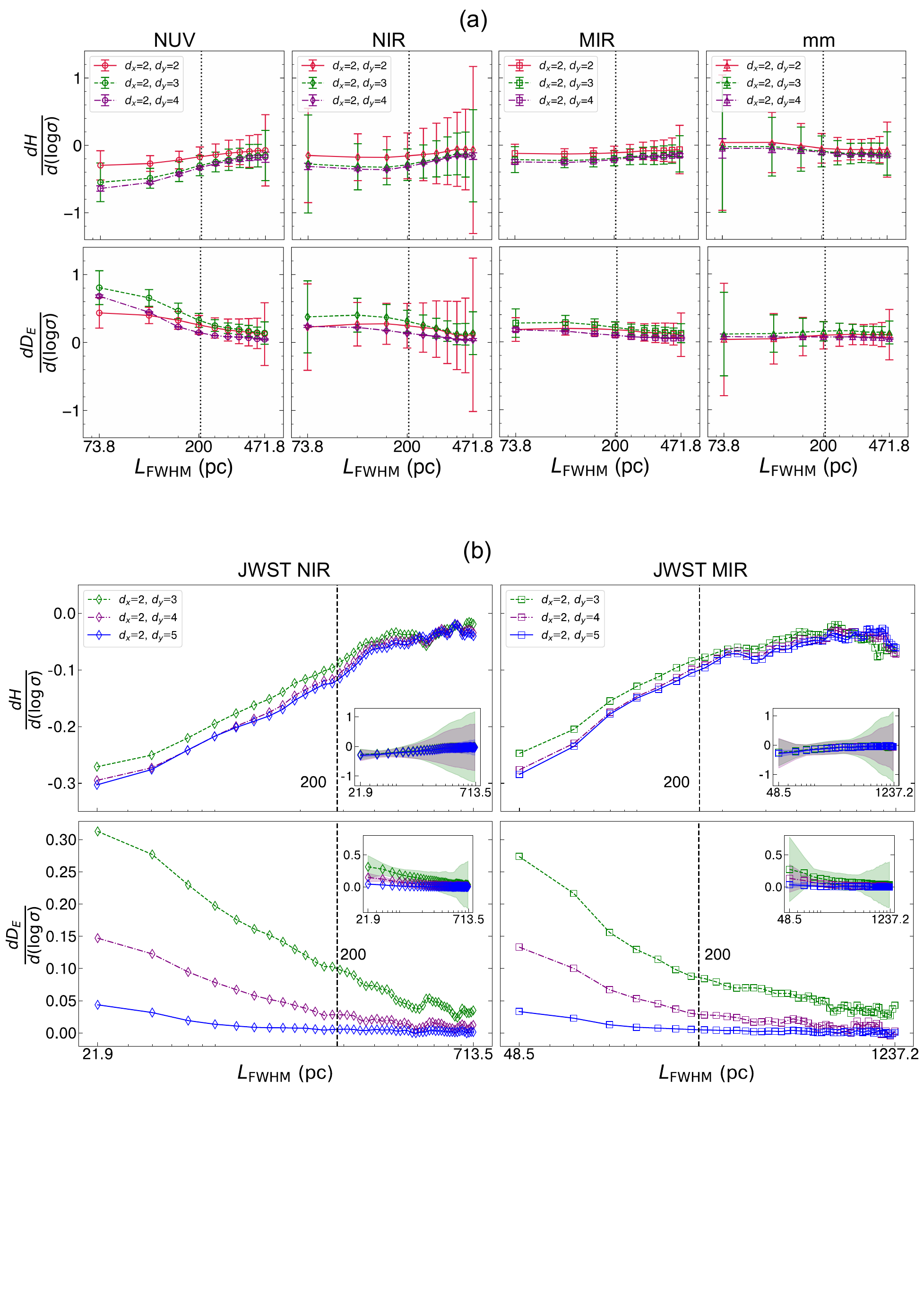}
\vspace{-3.5cm}
\caption{(a) $dH/d(\log \sigma)$ and $dD_E/d(\log \sigma)$ versus $L_{\rm FWHM}$ for the four images of NGC 628. The black dotted vertical line  marks $L_{\rm FWHM}^{(c)}=196$\,pc.  
(b) Similar plots as in (a) but for high-resolution JWST NIR and MIR images. Insets display the corresponding plots with error bars shown as colored shaded regions. Embedding dimensions are indicated in the legend.}
\label{fig:slopes_all4}
\end{figure*}

The value of $\sigma^{(c)}$ quoted in the previous subsections is inferred from the comparative behavior of $H$, $D_E$, and $C$ across wavelengths as a function of $\sigma$ as well as surrogate analysis. Since the scale dependence varies across individual wavelengths, we further analyze the slopes of $H$ and $D_E$ with respect to $\log\sigma$ for each wavelength, as shown versus $L_{\rm FWHM}$ in Figure~\ref{fig:slopes_all4}(a). These plots reveal significant variation among the wavelengths, indicating that the observed emissions arise from a mix of physical processes, each dominating at different spatial scales.

The original JWST NIRCam F360M and MIRI F2100W images have plate scales 3\,pc and 5.2\,pc, respectively, compared to the 19.89\,pc used previously. We recompute $H$ and $D_E$ using full-resolution data (see table~\ref{table1} for angular resolution) over a broader range of smoothing scales than in Figure~\ref{fig:slopes_all4}(a), enabling finer-scale fluctuation detection. With pixel number of the order of $\sim 10^6$, we now use $(d_x,d_y)=(2,3),(2,4)$ and $(2,5)$. The resulting slopes are shown in Figure~\ref{fig:slopes_all4}(b), with uncertainties indicated by the shaded regions in the insets.

While the plots in panel (b) are qualitatively similar to the corresponding NIR and MIR results in Figure~\ref{fig:slopes_all4}(a), they also reveal notable differences that reflect the added sensitivity to small-scale structures. Interestingly, in both NIR and MIR, the slopes of $H$ and $D_E$  flatten and show oscillations on scales larger than  200\,pc. This behavior at high $\sigma$ may reflect statistical fluctuations induced by large-scale galactic structures. Additionally, the associated uncertainties increase with $\sigma$, due to the reduced number of independent regions at larger smoothing scales. Important thing to be noticed is that the embedding dimensions cover a much smaller physical scale in Figure~\ref{fig:slopes_all4}(b) compared to Figure~\ref{fig:slopes_all4}(a). Hence, small embedding dimensions show more fluctuations at the same large smoothing scales. These fluctuations may also stem from variations in structure size with galactocentric distance, as some large features appear in the galaxy's outskirts. However, this study is limited to the inner disk of NGC 628.

\subsection{Ordinal Network Global Node entropy \texorpdfstring{S$_{\rm{gn}}$}{} across scales}
\label{sec:s5c}
\begin{figure}
\hspace{3.5cm} $d_x=2,\, d_y=3$\\
\hspace{-.5cm}\includegraphics[scale=0.6]{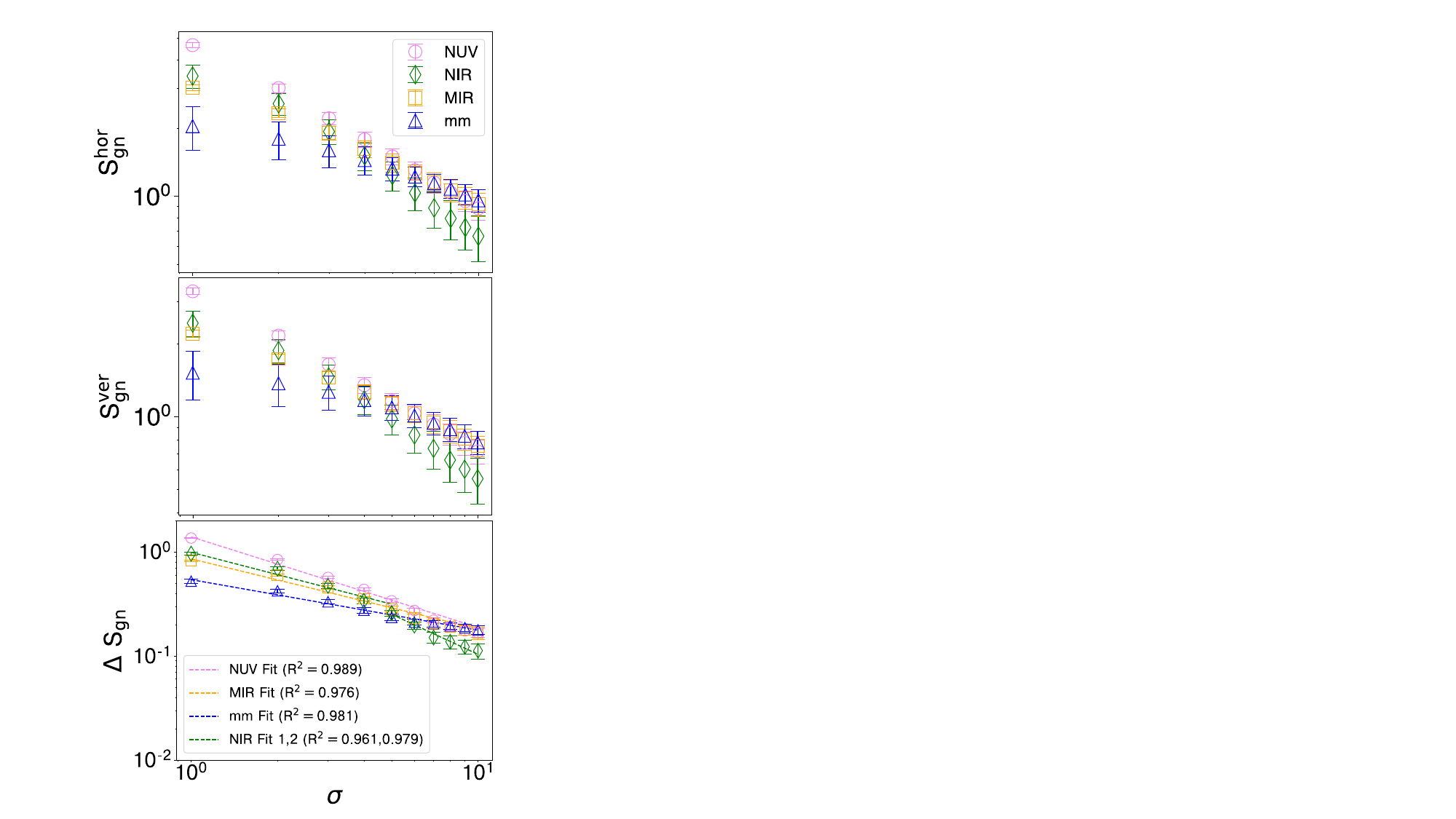}
\caption{Variation of global node entropy S$_{\rm gn}$ with smoothing scale $\sigma$  for embedding dimension $(d_x,d_y)=(2,3)$: Horizontal global node entropy, S$^{\rm hor}_{\rm gn}$ ({\em Top}), vertical global node entropy, S$^{\rm ver}_{\rm gn}$ ({\em Middle}), global node entropy difference, $\Delta$S$_{\rm gn}=$S$^{\rm hor}_{\rm gn}-$S$^{\rm ver}_{\rm gn}$, ({\em Bottom}) versus $\sigma$ on log-log plots. In the bottom panel, dashed lines represent inverse power-law fits ($\Delta$S$_{\rm gn}(\sigma)=a\sigma^{-b}$) with corresponding R$^2$-values for the goodness of fit (indicated in the legend). Values of power-law exponents $b$ are: 0.8705 (NUV), 0.6730 (MIR), and 0.4854 (mm). For NIR, we have fitted in two regimes with $b=0.7133$ $\&$ $1.2793$.}
\label{fig:7}
\end{figure}

We map the multiwavelength NGC 628 images at different $\sigma$ (in pixel units) into horizontal and vertical ordinal networks. We compute the corresponding horizontal (S$_{\rm gn}^{\rm hor}$) and vertical (S$_{\rm gn}^{\rm ver}$) global node entropies, and their difference ($\Delta$S$_{\rm gn}=$S$_{\rm gn}^{\rm hor}-$S$_{\rm gn}^{\rm ver}$) using $(d_x,d_y)=(2,3)$ (Figure~\ref{fig:7}). The decrease of S$_{\rm gn}^{\rm hor}$, S$_{\rm gn}^{\rm ver}$ and $\Delta$S$_{\rm gn}$ with $\sigma$ indicates progressive loss of the overall structural information of the galaxy field with smoothing. We again fit inverse power-law functions (dashed lines): $\Delta$S$_{\rm gn}(\sigma)=a\sigma^{-b}$ to the data (markers), with exponents: $b=$ 0.8705 (NUV),  
0.6730 (MIR), and 0.4854 (mm) [R$^2$-values indicated in the legend]. Entropy decays quickly in NUV (as the structures in this emission seem relatively even without strong clustering), slowly in mm (structures persist across scales with strong clustering), and moderately in MIR (some clustering exists, but not as pronounced as in mm). For NIR, two $\sigma$ regimes emerge: $b=0.7133$ for $\sigma < 5$, and $b=1.2793$ for $\sigma \geq 5$, suggesting a transition from a relatively uniform distribution of low-mass stars at small scales to central bulk emission at larger scales. This result from the ordinal network global node entropies also supports the critical transition scale $\sigma^{(c)}$ observed for NIR in the preceding subsection. 

The observed negative power-law decay of $\Delta$S$_{\rm gn}(\sigma)$ in NGC 628 indicates a scale-free reduction of structural information, pointing to a hierarchical organization~\citep{barabasi1999emergence} in the distribution of structures across scales. This hierarchical organization likely emerges from the self-organization~\citep{bak1988self} of various physical processes at multiple scales of the galaxy NGC 628, reflecting its multiscale complexity.

\section{Conclusions}
\label{sec:s5}
This study presents a novel approach to understanding the morphological complexity of the galaxy NGC 628 through the lens of statistical physics of complex systems. We demonstrate how the structural features of NGC 628 can be characterized within the framework of complexity analysis applied to empirical data. Using publicly available multi-wavelength imaging data, we systematically analyze NGC 628’s morphology across UV, NIR, MIR, and mm wavelengths, probing multiple physical scales. Our analysis employs the framework of ordinal patterns - permutation entropy, disequilibrium, statistical complexity, and ordinal network-based metrics. 
Taking into account a common field of view - set by the JWST MIR image - and the lowest resolution determined by the UVIT image - for the four images of NGC 628, our analysis probes smoothing scales ranging from 74 to 472 pc, corresponding to a physical extent of approximately 10 kpc. 

Our first key finding is the emergence of a characteristic scale near  200\,pc, where all structural complexity measures - permutation entropy, disequilibrium, statistical complexity, and global node entropy - show a clear transition. Surrogate field analysis (Figure~\ref{fig:surr}(b)) further confirms the emergence of this transition scale, marking a shift in the ISM from fluctuations caused by the turbulence acting at the smaller scale to more self-organized structures influenced by galaxy dynamics and global gravitational instability, consistently across all embedding dimensions. The  200\,pc scale aligns with the $l_{\rm corr}$, indicating a physical connection and also lends support to the robustness of our method. Our approach quantitatively captures the impact of star formation, traced by UV emission, in driving ISM turbulence and enhancing spatial randomness. This effect is weaker in colder, clumpier CO-dominated regions. Beyond  200\,pc, morphological differences across wavelengths become negligible.

Our second main conclusion is that the values of $C$ and $H$ for the four NGC 628 wavelengths trace a single smooth curve on the complexity-entropy plane, which asymptotes towards an `attractor' trajectory associated with isotropic Gaussian random fields. This suggests a statistically isotropic spatial distribution of the structures in NGC 628.

This pilot study demonstrates the effectiveness of ordinal patterns in probing the internal structure of galaxies and linking it to their non-equilibrium statistical states. A natural next step is to extend this analysis to a larger, morphologically diverse galaxy sample. Such a study could identify both distinguishing features for galaxy classification and underlying universal properties. This approach may also be combined with established geometrical and topological methods to uncover finer structural details, e.g., a topological analysis of HI in the Large Magellanic Cloud~\citep{kim2007topology} revealed four characteristic scales, each associated with distinct features, clumps, or holes, arising from different physical processes. While our ordinal pattern framework captures structural transition scales, it does not yet differentiate between topological types like holes or clumps or even structures from the star formation hierarchy~\citep{Gusev:2014}. Incorporating such distinctions is a direction we plan to pursue in future work.

Our findings may be viewed within a broader cosmological context. On sufficiently large scales, the cosmological density field is well approximated by a Gaussian random field, and its statistical properties are therefore fully described by the power spectrum. As gravitational instability drives nonlinear structure formation, mode coupling generates phase correlations, leading to coherent filamentary and clustered structures that cannot be captured by second-order statistics alone. Our results suggest that for galaxies, an opposite scale dependence may operate: at sufficiently small spatial scales, the surrogate and original galaxy images remain close in the complexity–entropy plane, indicating that local structure is largely consistent with the information encoded in the power spectrum. At scales around $\sigma=\sigma^{(c)}$, however, significant deviations emerge, marking the dominance of nonlinear, phase-dependent correlations. This behavior is consistent with a picture in which phase correlations are most significant in the nonlinear regime of structure formation in the universe, wherein both very large and very small scales exhibit statistics closer to Gaussian behavior.

\section*{Acknowledgements}  
\noindent We thank the referees for carefully reading the manuscript and giving valuable comments and suggestions. ALC acknowledges the APCTP (JRG program) through the Science and Technology Promotion Fund and Lottery Fund of the Korean Government and the Korean Local governments-Gyeongsangbuk-do Province and Pohang City. CBP is supported by KIAS Individual Grants (PG016903) at the Korea Institute for Advanced Study, and the National Research Foundation of Korea (NRF) grant funded by the Korean government (MSIT; RS-2024-00360385).

This publication uses data from the UVIT, which is part of the AstroSat mission of the Indian Space Research Organisation (ISRO), archived at the Indian Space Science Data Centre (ISSDC). We gratefully thank all the members of various teams for supporting the project from the early stages of design to launch and observations in orbit. 

  This work is based on observations made with the NASA/ESA/CSA JWST. The data were obtained from the Mikulski Archive for Space Telescopes at the Space Telescope Science Institute, which is operated by the Association of Universities for Research in Astronomy, Inc., under NASA contract NAS 5-03127. The observations are associated with the JWST program 2107. The specific JWST observation analyzed can be accessed
via DOI:\href{10.17909/q9f5-zy15}{10.17909/q9f5-zy15}. This paper makes use of the following ALMA data: ADS/JAO.ALMA\#2012.1.00650.S. ALMA is a partnership of ESO (representing its member states), NSF (USA), and NINS (Japan), together with NRC (Canada), MOST and ASIAA (Taiwan), and KASI (Republic of Korea), in cooperation with the Republic of Chile. The Joint ALMA Observatory is operated by ESO, AUI/NRAO, and NAOJ. The National Radio Astronomy Observatory is a facility of the National Science Foundation operated under a cooperative agreement by Associated Universities, Inc. 

\facilities{Astrosat (UVIT), JWST, ALMA}

\software{astropy \citep{2013A&A...558A..33A,2018AJ....156..123A}}

\appendix 
\section{Illustration of the construction of ordinal patterns}
\label{sec:append}

Let us consider a trivial example where the data  $Y$ is a $3\times 3$ matrix  given by,
\be
Y = \left( \ba{ccc} 4&9&5\\ 0&2&7\\ 2&6&5\ea \right).
\ee
For ease of writing, the data elements are taken to be integers. The actual galaxy data we analyze is real-valued. 

Let $d_x=2=d_y$ with unit embedding delay. Then $n_x=2,\ n_y=2$, which gives 4 sub-matrices. The steps of flattening each sub-matrix $y$ to a sequence $z$, mapping $z$ to a symbolic sequence $A$, sorting  $z$ to get $z_S$, and then obtaining the permuted symbolic sequence $A_S$, are explicitly demonstrated in the table below. 
\vskip .2cm
{\small{
\begin{tabular}{ccccc}
\hline
\ \texttt{Sub-matrix} & &  \texttt{Flatten and map to symbolic sequence}  & &  \quad \ \texttt{Sort and permute}\\
\hline
$y = \left( \ba{cc} 4&9\\ 0&2 \ea \right)$ & $\Rightarrow $ &\hskip 1cm $ \ba{cc} z =(4,9,0,2)\\ 
\teal{A= (0,1,2,3)} \ea $  & $\Rightarrow $ & \hskip .5cm $\ba{cc} z_S =(0,2,4,9)\\ \teal{A_S= (2,3,0,1)} \ea$ \\
\hline
$y = \left( \ba{cc} 9&5\\ 2&7 \ea \right)$ & $\Rightarrow $ &\hskip 1cm $ \ba{cc} z =(9,5,2,7)\\ 
\teal{A= (0,1,2,3)} \ea $  & $\Rightarrow $ & \hskip .5cm $\ba{cc} z_S =(2,5,7,9)\\ \teal{A_S= (2,1,3,0)} \ea$ \\
\hline
$y = \left( \ba{cc} 0&2\\ 2&6 \ea \right)$ & $\Rightarrow $ & \hskip 1cm$ \ba{cc} z =(0,2,2,6)\\ 
\teal{A= (0,1,2,3)} \ea $  & $\Rightarrow $ & \hskip .5cm $\ba{cc} z_S =(0,2,2,6)\\ \teal{A_S= (0,1,2,3)} \ea$ \\
\hline
$y = \left( \ba{cc} 2&7\\ 6&5 \ea \right)$ & $\Rightarrow $ & \hskip 1cm$ \ba{cc} z =(2,7,6,5)\\ 
\teal{A= (0,1,2,3)} \ea $  & $\Rightarrow $ & \hskip .5cm $\ba{cc} z_S =(2,5,6,7)\\ \teal{A_S= (0,3,2,1)} \ea$ \\
\hline
\end{tabular}
}}
\vskip .3cm
We get 4 permutation states (ordinal patterns), which are given by the symbolic sequences $A_S$. These constitute the ordinal sequence $\{\pi^j_i\}$. Note that $(d_xd_y)! = 4!$ gives 24 possible distinct basis states ($\psi_k$). The number of elements of $\{\pi^j_i\}$ is less than 24  because $Y$  is $3\times 3$, which violates the requirement  $d_x\ll N_x,\ d_y\ll N_y$ with $N_x=3=N_y$.   

\section{Motivation of Surrogate analysis}
\label{sec:appendb}
We now explain the statistical motivation of the surrogate analysis presented in section~\ref{sec:surr}. The constructed surrogate fields are GRFs that preserve the power spectrum of the galaxy while randomizing Fourier phases. Since the power spectrum determines the second-order statistics or two-point correlation function of a stationary random field, these surrogates retain all second-order statistical properties of the original galaxy image, including variance and correlation length. On the other hand, randomizing the phase removes higher-order dependencies in the data structure, yielding a GRF that is consistent with the same power spectrum as the original image. Any systematic deviation of the permutation entropy or statistical complexity from the surrogate ensemble, therefore, indicates the presence of higher-order spatial organization beyond that encoded in the power spectrum. Thus, constructing galaxy surrogates that preserve the exact galaxy power spectrum but have randomized Fourier phases would be an objectively appropriate procedure for detecting transition scales in a given galaxy image.

\bibliography{references}{}
\bibliographystyle{aasjournalv7}

\end{document}